\documentclass[aps,prl,reprint,superscriptaddress,twocolumn]{revtex4-2}

\usepackage{graphicx}
\usepackage{dcolumn}
\usepackage{bm}
\makeatletter

\begin{document}

\title{Percolative Pathway to Stripe Order in KTaO$_3$-Based Superconductivity}

\author{Zhihao Chen}
\thanks{These authors contributed equally to this work.}
\affiliation{Beijing National Laboratory for Condensed Matter Physics and Institute of Physics, Chinese Academy of Sciences, Beijing 100190, China}
\affiliation{School of Physical Sciences, University of Chinese Academy of Sciences, Beijing 100190, China}

\author{Chun Sum Brian Pang}
\thanks{These authors contributed equally to this work.}
\affiliation{Department of Physics and Astronomy, University of British Columbia, Vancouver, BC V6T 1Z1, Canada}
\affiliation{Quantum Matter Institute, University of British Columbia, Vancouver, BC V6T 1Z4, Canada}

\author{Meng Yang}
\thanks{These authors contributed equally to this work.}
\affiliation{Beijing National Laboratory for Condensed Matter Physics and Institute of Physics, Chinese Academy of Sciences, Beijing 100190, China}

\author{Yuxin Wang}
\affiliation{Beijing National Laboratory for Condensed Matter Physics and Institute of Physics, Chinese Academy of Sciences, Beijing 100190, China}

\author{Kun Jiang}
\affiliation{Beijing National Laboratory for Condensed Matter Physics and Institute of Physics, Chinese Academy of Sciences, Beijing 100190, China}
\affiliation{School of Physical Sciences, University of Chinese Academy of Sciences, Beijing 100190, China}

\author{Bruce A. Davidson}
\affiliation{Department of Physics and Astronomy, University of British Columbia, Vancouver, BC V6T 1Z1, Canada}
\affiliation{Quantum Matter Institute, University of British Columbia, Vancouver, BC V6T 1Z4, Canada}

\author{Ilya Elfimov}
\affiliation{Department of Physics and Astronomy, University of British Columbia, Vancouver, BC V6T 1Z1, Canada}
\affiliation{Quantum Matter Institute, University of British Columbia, Vancouver, BC V6T 1Z4, Canada}

\author{George A. Sawatzky}
\affiliation{Department of Physics and Astronomy, University of British Columbia, Vancouver, BC V6T 1Z1, Canada}
\affiliation{Quantum Matter Institute, University of British Columbia, Vancouver, BC V6T 1Z4, Canada}

\author{Andrea Damascelli}
\affiliation{Department of Physics and Astronomy, University of British Columbia, Vancouver, BC V6T 1Z1, Canada}
\affiliation{Quantum Matter Institute, University of British Columbia, Vancouver, BC V6T 1Z4, Canada}

\author{Ke Zou}
\email{kzou@phas.ubc.ca}
\affiliation{Department of Physics and Astronomy, University of British Columbia, Vancouver, BC V6T 1Z1, Canada}
\affiliation{Quantum Matter Institute, University of British Columbia, Vancouver, BC V6T 1Z4, Canada}

\author{Zhi Gang Cheng}
\email{zgcheng@iphy.ac.cn}
\affiliation{Beijing National Laboratory for Condensed Matter Physics and Institute of Physics, Chinese Academy of Sciences, Beijing 100190, China}
\affiliation{School of Physical Sciences, University of Chinese Academy of Sciences, Beijing 100190, China}

\date{\today}

\begin{abstract}
The sensitivity of low-dimensional superconductors to fluctuations gives rise to emergent behaviors beyond the conventional Bardeen–Cooper–Schrieffer framework. Anisotropy is one such manifestation, often linked to spatially modulated electronic states and unconventional pairing mechanisms. Pronounced in-plane anisotropy recently reported at KTaO$_3$-based oxide interfaces points to the emergence of a stripe order in superconducting phase, yet its microscopic origin and formation pathway remain unresolved. Here, we show that controlled interfacial disorder in MgO/KTaO$_3$(111) heterostructures drives a percolative evolution from localized Cooper-pair islands to superconducting puddles and eventually to stripes. The extracted stripe width matches the spin precession length, suggesting a self-organized modulation governed by spin–orbit coupling and lattice-symmetry breaking. These findings identify disorder as both a tuning parameter and a diagnostic probe for emergent superconductivity in two-dimensional quantum materials.
\end{abstract}

\maketitle

Two-dimensional (2D) superconductivity is inherently intriguing because it departs in profound ways from its three-dimensional (3D) counterpart. As the dimension is reduced, true long-range order is prohibited at finite temperature. Phase coherence instead emerges through the Berezinskii–Kosterlitz–Thouless (BKT) mechanism \cite{berezinski1971,Kosterlitz1972,Kosterlitz1973} where vortices' mobility governs the transition between a dissipative state and a coherent superconducting state. In addition, the reduced dimensionality greatly amplifies the effects of extrinsically induced fluctuations and makes the superconducting state highly sensitive and tunable to external stimuli such as electric field, magnetic field, strain, and controlled disorder \cite{Kivelson2003,Haviland1989a}, making the 2D electron gas (2DEG) heavily influenced by its microscopic environment.

A particularly interesting platform for exploring fluctuation-governed superconductivity is provided by oxide interfaces that host 2DEGs within electrically correlated and structurally flexible perovskite lattices. The LaAlO$_3$/SrTiO$_3$ (LAO/STO) interface has played a central role in demonstrating how interfacial confinement, orbital multiplicity, soft phonon modes, and inversion-symmetry breaking collectively shape superconductivity \cite{Ohtomo2004,reyren2007}. The couplings between charge, spin, orbital, and lattice degrees of freedom endow 2DEGs with a rich landscape of exotic behaviors \cite{Brinkman2007,Li2011,Caviglia2008,Caviglia2010,Caviglia2010a}, making them ideal for exploring fluctuation-driven superconducting phases.

KTaO$_3$ (KTO)-based interfaces further enrich this landscape by its much stronger spin–orbit coupling (SOC) from Ta's 5d electrons and incipient polarization from the quantum paraelectric background \cite{Esswein2022,Fujishita2016}. These characteristics give rise to superconductivity that is unusually sensitive to fluctuations and symmetry considerations \cite{Liu2021,Chen2021a,Liu2023,Chen2021b}. One particularly striking feature, reported across several KTO-based heterostructures but absent in STO counterparts, is the pronounced in-plane anisotropy in superconducting transport \cite{Liu2021,Hua2024,Arnault2023a} although its microscopic origin remains unsettled. Various possibilities have been discussed, including anisotropic electronic responses enabled by strong SOC \cite{Arnault2023a,Bruno2019a}, magnetic proximitization \cite{Hua2024}, and parity-mixed superconducting pairing \cite{Zhai2025,Zhang2023}. The anisotropy motivates the proposal that a stripe-like superconducting texture is developed \cite{Liu2021,Hua2024}, yet its formation pathway remains elusive.

Here we use disorder as a tool to uncover the formation pathway of superconducting anisotropy in MgO/KTO(111) heterostructures. Although usually treated as distructive, disorder remarkably broadens the superconducting transition, allowing us to track the progression of spatial coherence from preformed Cooper pair islands to superconducting puddles and ultimately to stripes. The extracted stripe width matches the spin precession length of electrons, highlighting the key role that SOC plays in the stripe formation. Our study uncovers the correlation between superconducting fluctuations and emergent percolative order, establishing disorder as a universal tuning parameter in 2D superconductors with SOC. This mechanism may be broadly applicable to other low-dimensional superconductors.

\begin{figure}[t]
    \centering
    \includegraphics[width=1\linewidth]{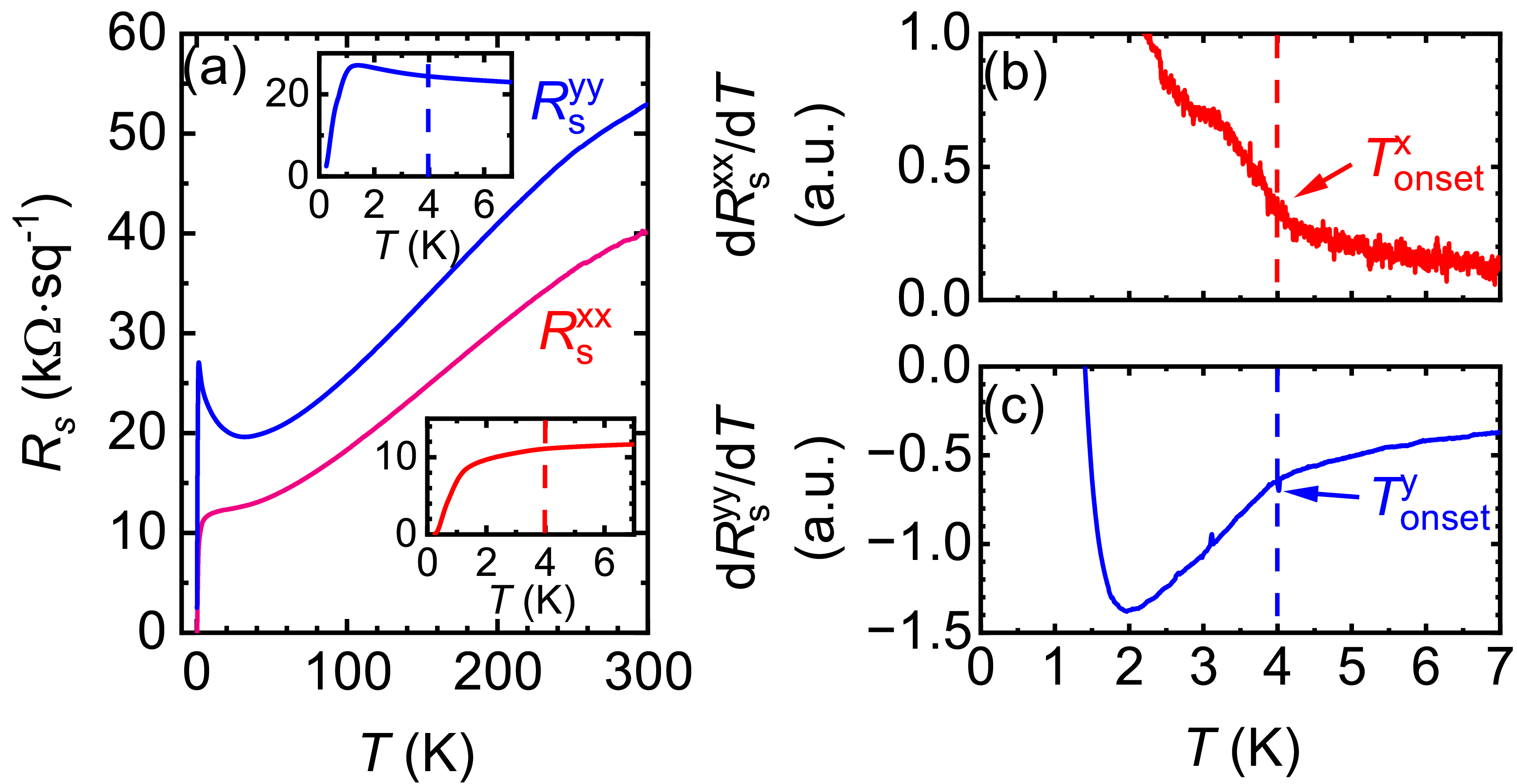}
    \caption{
        (a) Temperature dependence of sheet resistances $R_{s}^{xx}$ and $R_{s}^{yy}$ from 300 K to 0.25 K. The insets are zoom-in views of $R_{s}^{xx}$ and $R_{s}^{yy}$ below 7 K.
        (b, c) $dR_{s}^{xx}/dT$ and $dR_{s}^{yy}/dT$ below 7 K to display the kinks that signal the superconducting transition onset. 
        Normal state values for $R_{s}^{xx}$ and $R_{s}^{yy}$ are read at 4 K, denoted as $R_{N}^{x}$ and $R_{N}^{y}$, respectively. 
    }
    \label{fig1}
\end{figure}

The MgO/KTO heterostructures were prepared by molecular beam epitaxy where a reduction of the KTO surface generated a 2DEG with controlled interfacial disorder. Two samples were prepared -- data from one (S1) are presented in the main text and the other (S2) in the Supplemental Materials (SM) \cite{Suppl}. Growth details are provided in the SM  \cite{Suppl} and a separate publication \cite{Pang2025}.

We define the in-plane crystallographic directions $\lbrack 11\overline{2}\rbrack$ and $\lbrack 1\overline{1}0\rbrack$ as the $x$ and $y$, and the normal direction $[111]$ as $z$. Transport results reveal a pronounced in-plane anisotropy in both normal and superconducting states. Figure \ref{fig1}(a) shows that the sheet resistivity along $x$ ($R_{s}^{xx}$) remains metallic but a metal-insulator transition (MIT) emerges below 50 K in $y$ as an upturn in $R_{s}^{yy}$. The MIT was not observed in previous KTO-based heterostructures, underscoring its origin from disorder. Kinks in $dR_{s}/dT$ are observed in both directions at 4 K (Figs. \ref{fig1}(b, c)) which signal the superconducting onset. However, unlike the regular decrease in $R_{s}^{xx}$, the anomalous rise in $R_{s}^{yy}$ suggests the failure of developing macroscopic coherence in $y$ until 1 K, below which rapid decreases are observed in both directions. Notably, the transition is significantly broadened. Instead of reaching zero-resistance at $\sim1$ K as in previous studies \cite{Liu2021,Chen2021a}, it is incomplete even at the base temperature of our measurements at 0.25 K.

\begin{figure}[b]
    \centering
    \includegraphics[width=1\linewidth]{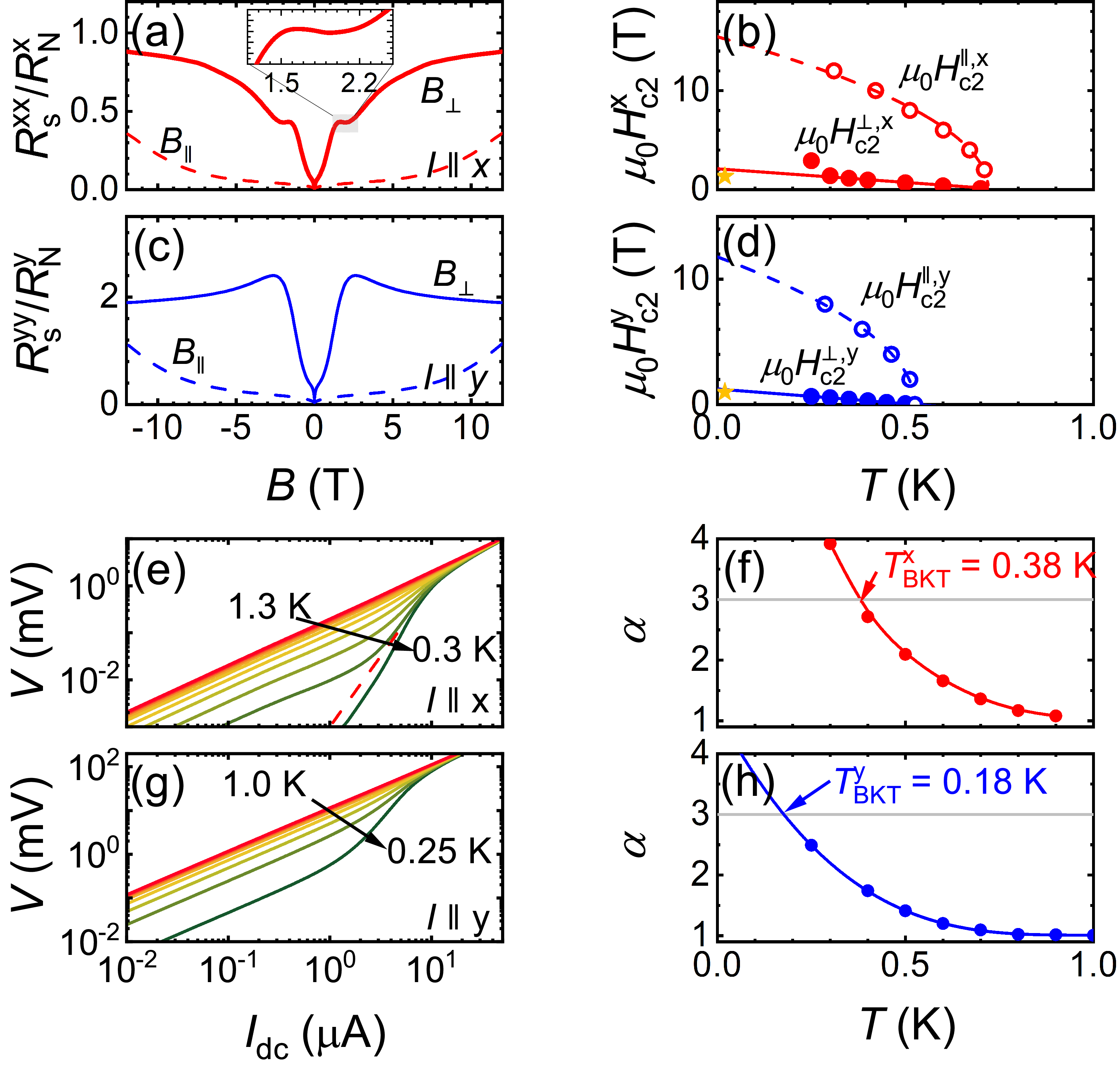}
    \caption{
        (a) Normalized MR measured at 0.25 K with $I \parallel x$. The inset is a zoom-in view of the negative MR anomaly between 1.2 T and 2.5 T. 
        (b) $\mu_0H_{c2}$ vs. $T$ measured with $I \parallel x$. The values are read when $R_s=0.5R_N^x$, and the yellow star marks the Pauli limit. 
        Data sets measured with $I \parallel y$ are plotted in Panels (c, d) in the same order as (a, b). 
        (e) $V$-$I$ curves measured with $I \parallel x$. The red dashed line plots the scaling law $V \propto I^{\alpha}$ with $\alpha=3$. 
        (f) $\alpha$ vs. $T$. The grey horinzontal line marks $\alpha=3$. 
        The $V$-$I$ data measured with $I \parallel y$ are plotted in Panels (g, h) in the same order as (e, f).
    }
    \label{fig2}
\end{figure}

The nature of 2D superconductivity is confirmed by magnetoresistance (MR) and $V$-$I$ measurements as shown in Fig. \ref{fig2}. By fitting the critical magnetic field $\mu_0H_{c2}$ with Tinkham's model \cite{Tinkham1963} as shown in Figs. \ref{fig2}(b, d), we extract an effective thickness of the superconducting layer $d_{t} = 5.93$ nm but direction-dependent Ginzburg-Landau coherence lengths in the 0 K limit: $\xi_{GL}^{x}(0) = 12.49$ nm and $\xi_{GL}^{y}(0) = 16.29$ nm. Notably, the in-plane critical fields significantly exceed the Pauli limit, indicating a significant role of SOC to enhance superconductivity. In addition, a slightly negative MR anomaly is resolved under perpendicular field between 1.5 T and 2.2 T, marking a vortex reconfiguration due to spatial restriction \cite{Jiang2020,Zhang2021a}. Its appearance only with $I \parallel x$  anticipates the emergence of stripe order despite the nominal 2D geometry as discussed below.

DC $V$-$I$ measurements identifies the BKT transition at $T_{BKT}=0.38$ K. It is measured with $I \parallel x$ as shown in Fig. 2(e, f), where $T_{BKT}$ marks the localization of vortices upon pairing. However, $T_{BKT}$ is much lower for $I \parallel y$,  estimated near $0.18$ K as in Figs. \ref{fig2}(g, h)), suggesting the strong directional dependence of vortex motion.

\begin{figure}[t]
    \centering
    \includegraphics[width=1\linewidth]{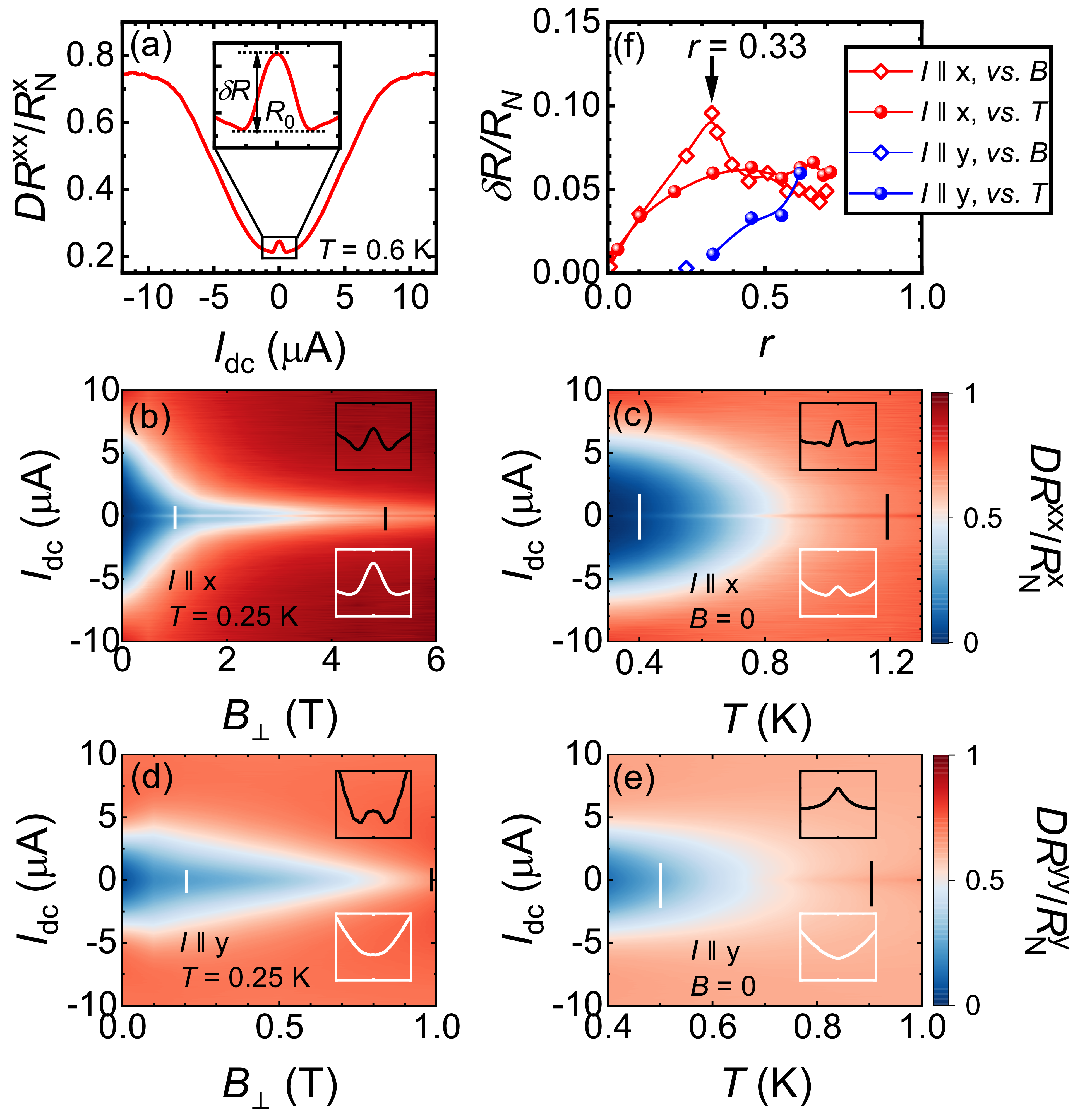}
    \caption{
    (a) Normalized $DR$ measured with $I \parallel x$ at 0.6 K. A ZBRP is resolved near $I_{dc}=0$ as shown in the inset. The peak height is denoted as $\delta R$ and the baseline is $R_{0}$. 
    (b-e) Normalized $DR$ measured at various magnetic fields and temperatures. The insets show $DR$ vs. $I_{dc}$ along the trajectories in the same colors.
    (f) Dependence of peak height $\delta R/R_N$ on $r$.  The arrow marks the sharp drop of $\delta R$ at {r = 0.33}, equivalently with $B = 1.5$ T.
    }
    \label{fig3}
\end{figure}

We further probe the vortex motion by applying DC current bias and measure differential resistance (DR). A current bias ($I_{dc}$) would trigger vortice's transverse motion via Lorentz force, thus leading to finite dissipation. Figure \ref{fig3}(a) exhibits a typical DR dependence on $I_{dc}$, which shows an additional resistance peak near $I_{dc}=0$ in addition to the regular superconducting transition as $I_{dc}$ increases. Such a zero bias resistance peak (ZBRP) can always be resolved at multiple magnetic field or temperature for $I \parallel x$, but only appears above 0.9 T or 0.6 K for $I \parallel y$ (see Fig. \ref{fig3} and Fig. S2 in SM \cite{Suppl}). DR is usually expected to rise with $I_{dc}$ in 2D situation \cite{Tinkham2015a}. However, the peak suggests a suppression of vortex motion as $I_{dc}$ is initially applied.  Although it is anomalous for 2D framework, the ZBRP has been observed in quasi-1D superconductors such as stripes and nanowires \cite{Zhang2021,Xiong1997,Berdiyorov2012} and ascribed to vortex pinning at boundaries as illustrated in Figs. S7(a, b). The peak observed in our study thus strongly suggests spatially heterogeneity at the MgO/KTO interface. More interestingly, it exhibits a clear anisotropy: vortex motion is strongly confined in $y$ as probed by $I \parallel x$, but the confinement in $x$ is  lifted  deep in the superconducting state as revealed by its gradual disappearance with $I \parallel y$. These observations point to a percolative evolution from isolated superconducting puddles to coherent stripes.

To characterize the transition quantitatively, we extract the peak height $\delta R$ and the background of its emergence $R_{0}$ as denoted in Fig. \ref{fig3}(a), both measured with $I \parallel x$. The ratio $r = R_{0}/R_{N}^{x}(4\ \mathrm{K})$ is used as an indicator for the progression of superconducting transition where $R_{N}^{x}(4\ \mathrm{K})$ is the sheet resistance of normal state at 4 K. As evidenced in Fig. \ref{fig3}(f), $\delta R$ first increases with $B$ but then drastically drops at 1.5 T, equivalently at $r = 0.33$. In contrast, $\delta R$ grows smoothly with $T$ and joins its field dependence at $r = 0.4$. We further compare the peak height measured with $I \parallel y$ (with the definition of $r$ explained in SM \cite{Suppl} for this case). Remarkably, its appearances either at 0.6 K or 0.9 T are also near $r = 0.33$. Furthermore, the negative MR anomaly between 1.5 T and 2.2 T shown in Fig. \ref{fig2}(a) takes place as $r = 0.33 \sim 0.38$. These coincidental features collectively confirm the coalescence of superconducting puddles into stripes.

\begin{figure*}[]
    \centering
    \includegraphics[width=1\linewidth]{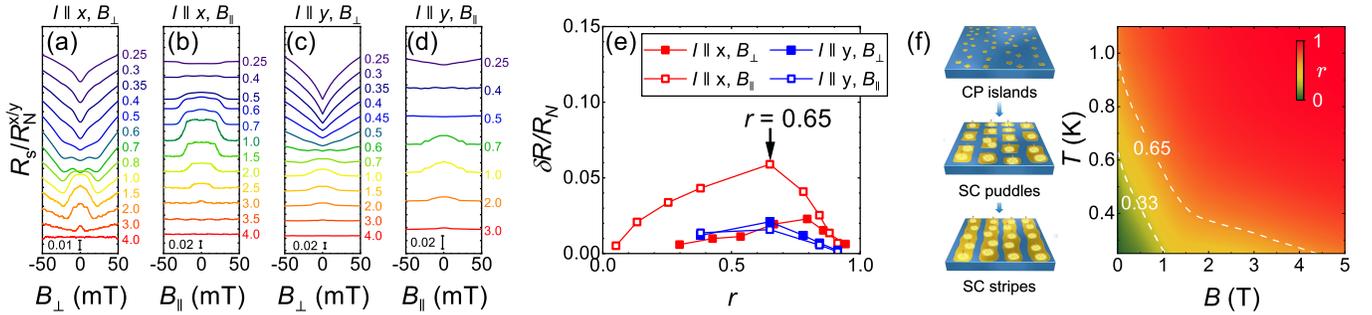}
    \caption{
    (a-d) Normalized MR measured with various combinations of current directions and magnetic field orientations as labelled above each panel. Data curves are vertically shifted with measurement temperatures labelled next to each curve. 
    (e) Dependence of normalized peak height $\delta R/R_N$ on $r$. The arrow marks $r = 0.65$ beyond which $\delta R/R_N$ gradually decays. 
    (f) Phase diagram of the superconducting transition with the isocurves $r = 0.65$ and $r=0.33$ marking the crossovers from Cooper pair islands to superconducting puddles and to stripes.
    }
    \label{fig4}
\end{figure*}

To advance the investigation, we conducted MR measurements to probe the vortex motion in magnetic field. In addition to Meissner effect displayed in Figs. S3,  an MR peak can be identified at $B = 0\ $ in all measurements as shown in Figs. \ref{fig4}(a-d). The peak appears below 4 K and broadens upon cooling, reaching $\pm30$ mT before decaying. We exclude weak localization (WL) as its origin because it is observed under parallel field. Furthermore, we rule out ferromagnetic origin by carefully sweeping magnetic field back and forth but observing no hysteresis (Fig. S4). Referring to the stripe order inferred by the DR results, we ascribe the MR peaks to the interaction between vortices and Meissner currents in quasi-1D systems \cite{Zhang2021,Cordoba2013}. External magnetic field induces screening current near stripe edges, resulting the Gibbs free energy to rise near edges. It develops Bean-Livingston barriers \cite{Bean1964} sandwiching a minimum at the stripe center (Fig. S6(b)). Vortex motion is therefore suppressed (Figs. S7(c, d)), leading to a resistance drop. More significantly, the MR peak exhibits directional sensitivity as well. It persists down to 0.4 K for $I \parallel x$ (equivalently $r \approx 0.05$, see Fig. \ref{fig4}(b)) but is only visible above 0.6 K for $I \parallel y$ with reduced heights ($r > 0.33$, see Fig. \ref{fig4}(d)). Such anisotropy aligns with the DR peak and provides an independent signature of stripe formation from puddle coalescence. It is worth noting the transition from resistance peaks to dips as cooling in perpendicular field (see Figs. \ref{fig4}(a, c)). The dips are contributed by weak anti-localization (WAL) which bring difficulties for reliable analysis on the peaks.

Referred to the analogues in quasi-1D systems \cite{Zhang2021,Cordoba2013,Rogachev2006,Stan2004}, we can estimate the stripe width by the peak width. $B_{V}$ marks the field at which the central minimum of Gibbs free energy is as low as the stripe edges. It is associated with the stripe width $w$ and superconducting coherence length $\xi$: $B_{V} = \frac{2\Phi_{0}}{\pi w^{2}}\ln(\frac{w}{4\xi_{GL}})$, where $\Phi_{0} = h/2e = 2.07 \times 10^{- 15}$ Wb is magnetic flux quanutm \cite{Likharev1971,Maksimova1998}.  Vortex dissipation is minimized for $B=B_V$ while any larger field would lower the minimum further and introduce extra dissipation by vortex tunnelling through the barriers and entering the stripe. By taking $B_{V} = 30$ mT and $\xi_{GL}(0.6\ \mathrm{K}) = 17.7$ nm, we estimate $w \approx 83\ $nm. The ratio $w/\xi_{GL} \approx 4.7$ validates the stripe geometry.

Evidenced by the simultaneous decreases in $R_s^{xx}$ and $R_s^{yy}$, phase coherence develops below 1 K. Conversely, it is gradually lost at higher temperatures and quasiparticle excitations are activated to dominate the dissipation. They effectively fragment the superconducting network and isolate the puddles into localized Cooper-pair islands. Such degradation gives rise to the diminishing MR peaks for $T>1$ K (equivalently $r > 0.65$, see Fig. \ref{fig4}(e)).

Our measurements establish a consistent picture of superconductivity development in the MgO/KTO interface as illustrated in Fig. \ref{fig4}(f). Cooper-pair islands set in at 4 K, progressively expand into superconducting puddles, and coalesce into stripes below 0.6 K. This pathway is supported by the directional sensitivity of ZBRPs in both DR and MR results as signatures of vortex confinement. A similar progression can be also realized by tuning magnetic field -- a reduction in magnetic field promotes Cooper-pair islands to expand and coalesce into stripes. Notably, these transitions are consistently governed by the indicator $r$ which marks the crossovers at $r = 0.65$ and $r = 0.33$, irrespective of whether the driving force is temperature or magnetic field.

The stripe order reflects a spatial symmetry breaking. Unlike the stripe phases observed in underdoped cuprates which are contributed by spin and charge orders at atomic length scales \cite{Berg2007,Kivelson2003,Ruan2018,Ye2023,Wollny2009,Wilson2006,Moshchalkov2000a,Moshchalkov2000b,Li2007}, the stripe structures in KTO extend over 80 nm -- comparable to the spin precession lengths of itinerant electrons revealed by WAL effect \cite{Silotia2024,Nakamura2009,Zou2022}. It suggests the key role of SOC in stabilizing the stripe order. Interestingly, the MIT illustrated by $R_s^{yy}$  emerges near 50 K (Fig. \ref{fig1}(a)). Despite preceding the superconducting onset at 4 K, it coincides with the emergence of recently-reported ferroelectricity \cite{Zhang2025}. Furthermore, $R_s^{yy}$ is minimized near 35 K during the MIT, matching the appearance of polar nanoregions in doped KTO \cite{Trybula2015}. Although macroscopic ferroelectricity is suppressed in quantum paraelectric KTO \cite{Esswein2022,Fujishita2016}, defects such as oxygen vacancies and anti-sites introduce strain to stabilize local polarization \cite{Trybula2015,Tyunina2010,Junquera2023,Kleemann2011}. These locally polarized regions are enriched with Rashba-like splittings with reduced symmetry \cite{Venditti2023} in addition to the $C_{3}$-symmetric Rashba SOC imprinted by the (111) plane \cite{Arnault2023a,Bruno2019a}. Such splittings are expected to enhance electron-phonon coupling along a preferred direction \cite{Gastiasoro2023} and promotes anisotropy in superconducting ground state. 


Disorder has been successfully used to probe the transition sequence of SrTiO$_3$-based superconductivity \cite{Chen2018}. In our study, disorder broadens the superconducting transition and make intermediate phases experimentally accessible. The wide transition window from 4 K to 0.25 K allows us to resolve the distinct stages, which would otherwise merge into a sharp transition in less disordered interfaces. Conversely, the transition of another more disordered sample S2 (characterized as shown in SM \cite{Suppl}) stops at the puddle phase without forming stripes (Figs. S8-S10),  confirming further disorder's tuning capability. We are aware of the similar zero-bias MR peaks reported in a Hall device with patterned array of insulating islands on LaAlO$_3$/KTaO$_3$(110) interfaces \cite{Wang2025}. Vortices are restricted by the artificially imposed boundaries in this case, distinct from the self-organized superconducting textures in our study. These comparisons highlight the versatility of oxide interfaces and the effectiveness of disorder engineering to reveal correlated electronic states. It is potentially feasible to use disorder as an effective parameter in tuning not only the strength but also the percolation of superconductivity in reduced dimensions.

In summary, we successfully demonstrate the percolative pathway for a superconducting stripe order to develop at MgO/KTaO$_3$(111) interfaces by introducing controlled disorder. This approach broadens the superconducting transition and unveils a continuous evolution from localized Cooper-pair islands to percolative superconducting puddles, and ultimately to stripes. The extracted stripe width is comparable to electrons' spin-precession length, implicating spin–orbit coupling as a key factor for stripes to form. These findings underscore the critical interplay among spin–orbit coupling, disorder, and the lattice symmetry of the quantum paraelectric host in defining the superconducting ground state. This mechanism could be broadly relevant to other low-dimensional superconductors where strong spin–orbit interaction and local symmetry breaking coexist, such as superconducting films and transition-metal dichalcogenides in addition to oxide interfaces. Our work establishes a versatile strategy to probe and engineer emergent electronic phases in reduced dimensions through controlled disorder and interfacial design.

\section{acknowledgments}
We appreciate discussions and assistance from Hongming Weng, Yi Zhou, Zhenyu Li, Haiwen Liu, and Alexander Balatsky. This research was supported by the National Natural Science Foundation of China (Grant No. T2325026), the National Key R\&D Program of China (Grant No. 2021YFA1401902), the Key Research Program of Frontier Sciences, CAS (Grant No. ZDBS-LY-SLH0010), and the CAS Project for Young Scientists in Basic Research (Grant No. YSBR047). This research was undertaken thanks in part to funding from the Max Planck–UBC–UTokyo Centre for Quantum Materials, the Canada First Research Excellence Fund, Quantum Materials and Future Technologies, the Natural Sciences and Engineering Research Council of Canada (NSERC), Canada Foundation for Innovation (CFI), the Department of National Defence (DND); the British Columbia Knowledge Development Fund (BCKDF), the Canada Research Chairs Program (A.D.), and the CIFAR Quantum Materials Program (A.D.).

\bibliographystyle{apsrev4-2}
\bibliography{prl_main.bib}

\begin{thebibliography}{58}%
\makeatletter
\providecommand \@ifxundefined [1]{%
 \@ifx{#1\undefined}
}%
\providecommand \@ifnum [1]{%
 \ifnum #1\expandafter \@firstoftwo
 \else \expandafter \@secondoftwo
 \fi
}%
\providecommand \@ifx [1]{%
 \ifx #1\expandafter \@firstoftwo
 \else \expandafter \@secondoftwo
 \fi
}%
\providecommand \natexlab [1]{#1}%
\providecommand \enquote  [1]{``#1''}%
\providecommand \bibnamefont  [1]{#1}%
\providecommand \bibfnamefont [1]{#1}%
\providecommand \citenamefont [1]{#1}%
\providecommand \href@noop [0]{\@secondoftwo}%
\providecommand \href [0]{\begingroup \@sanitize@url \@href}%
\providecommand \@href[1]{\@@startlink{#1}\@@href}%
\providecommand \@@href[1]{\endgroup#1\@@endlink}%
\providecommand \@sanitize@url [0]{\catcode `\\12\catcode `\$12\catcode
  `\&12\catcode `\#12\catcode `\^12\catcode `\_12\catcode `\%12\relax}%
\providecommand \@@startlink[1]{}%
\providecommand \@@endlink[0]{}%
\providecommand \url  [0]{\begingroup\@sanitize@url \@url }%
\providecommand \@url [1]{\endgroup\@href {#1}{\urlprefix }}%
\providecommand \urlprefix  [0]{URL }%
\providecommand \Eprint [0]{\href }%
\providecommand \doibase [0]{https://doi.org/}%
\providecommand \selectlanguage [0]{\@gobble}%
\providecommand \bibinfo  [0]{\@secondoftwo}%
\providecommand \bibfield  [0]{\@secondoftwo}%
\providecommand \translation [1]{[#1]}%
\providecommand \BibitemOpen [0]{}%
\providecommand \bibitemStop [0]{}%
\providecommand \bibitemNoStop [0]{.\EOS\space}%
\providecommand \EOS [0]{\spacefactor3000\relax}%
\providecommand \BibitemShut  [1]{\csname bibitem#1\endcsname}%
\let\auto@bib@innerbib\@empty
\bibitem [{\citenamefont {BEREZINSKI}(1971)}]{berezinski1971}%
  \BibitemOpen
  \bibfield  {author} {\bibinfo {author} {\bibfnamefont {V.~L.}\ \bibnamefont
  {BEREZINSKI}},\ }\href@noop {} {\bibfield  {journal} {\bibinfo  {journal}
  {Sov. Phys. JETP}\ }\textbf {\bibinfo {volume} {32}},\ \bibinfo {pages} {493}
  (\bibinfo {year} {1971})}\BibitemShut {NoStop}%
\bibitem [{\citenamefont {Kosterlitz}\ and\ \citenamefont
  {Thouless}(1972)}]{Kosterlitz1972}%
  \BibitemOpen
  \bibfield  {author} {\bibinfo {author} {\bibfnamefont {J.~M.}\ \bibnamefont
  {Kosterlitz}}\ and\ \bibinfo {author} {\bibfnamefont {D.~J.}\ \bibnamefont
  {Thouless}},\ }\href {https://doi.org/10.1088/0022-3719/5/11/002} {\bibfield
  {journal} {\bibinfo  {journal} {Journal of Physics C: Solid State Physics}\
  }\textbf {\bibinfo {volume} {5}},\ \bibinfo {pages} {L124} (\bibinfo {year}
  {1972})}\BibitemShut {NoStop}%
\bibitem [{\citenamefont {Kosterlitz}\ and\ \citenamefont
  {Thouless}(1973)}]{Kosterlitz1973}%
  \BibitemOpen
  \bibfield  {author} {\bibinfo {author} {\bibfnamefont {J.~M.}\ \bibnamefont
  {Kosterlitz}}\ and\ \bibinfo {author} {\bibfnamefont {D.~J.}\ \bibnamefont
  {Thouless}},\ }\href {https://doi.org/10.1088/0022-3719/6/7/010} {\bibfield
  {journal} {\bibinfo  {journal} {Journal of Physics C: Solid State Physics}\
  }\textbf {\bibinfo {volume} {6}},\ \bibinfo {pages} {1181} (\bibinfo {year}
  {1973})}\BibitemShut {NoStop}%
\bibitem [{\citenamefont {Kivelson}\ \emph {et~al.}(2003)\citenamefont
  {Kivelson}, \citenamefont {Bindloss}, \citenamefont {Fradkin}, \citenamefont
  {Oganesyan}, \citenamefont {Tranquada}, \citenamefont {Kapitulnik},\ and\
  \citenamefont {Howald}}]{Kivelson2003}%
  \BibitemOpen
  \bibfield  {author} {\bibinfo {author} {\bibfnamefont {S.~A.}\ \bibnamefont
  {Kivelson}}, \bibinfo {author} {\bibfnamefont {I.~P.}\ \bibnamefont
  {Bindloss}}, \bibinfo {author} {\bibfnamefont {E.}~\bibnamefont {Fradkin}},
  \bibinfo {author} {\bibfnamefont {V.}~\bibnamefont {Oganesyan}}, \bibinfo
  {author} {\bibfnamefont {J.~M.}\ \bibnamefont {Tranquada}}, \bibinfo {author}
  {\bibfnamefont {A.}~\bibnamefont {Kapitulnik}},\ and\ \bibinfo {author}
  {\bibfnamefont {C.}~\bibnamefont {Howald}},\ }\href@noop {} {\bibfield
  {journal} {\bibinfo  {journal} {Rev. Mod. Phys.}\ }\textbf {\bibinfo {volume}
  {75}},\ \bibinfo {pages} {1201} (\bibinfo {year} {2003})}\BibitemShut
  {NoStop}%
\bibitem [{\citenamefont {Haviland}\ \emph {et~al.}(1989)\citenamefont
  {Haviland}, \citenamefont {Liu},\ and\ \citenamefont
  {Goldman}}]{Haviland1989a}%
  \BibitemOpen
  \bibfield  {author} {\bibinfo {author} {\bibfnamefont {D.~B.}\ \bibnamefont
  {Haviland}}, \bibinfo {author} {\bibfnamefont {Y.}~\bibnamefont {Liu}},\ and\
  \bibinfo {author} {\bibfnamefont {A.~M.}\ \bibnamefont {Goldman}},\ }\href
  {https://doi.org/10.1103/PhysRevLett.62.2180} {\bibfield  {journal} {\bibinfo
   {journal} {Phys. Rev. Lett.}\ }\textbf {\bibinfo {volume} {62}},\ \bibinfo
  {pages} {2180} (\bibinfo {year} {1989})}\BibitemShut {NoStop}%
\bibitem [{\citenamefont {Ohtomo}\ and\ \citenamefont
  {Hwang}(2004)}]{Ohtomo2004}%
  \BibitemOpen
  \bibfield  {author} {\bibinfo {author} {\bibfnamefont {A.}~\bibnamefont
  {Ohtomo}}\ and\ \bibinfo {author} {\bibfnamefont {H.~Y.}\ \bibnamefont
  {Hwang}},\ }\href {https://doi.org/10.1038/nature02308} {\bibfield  {journal}
  {\bibinfo  {journal} {Nature}\ }\textbf {\bibinfo {volume} {427}},\ \bibinfo
  {pages} {423} (\bibinfo {year} {2004})}\BibitemShut {NoStop}%
\bibitem [{\citenamefont {Reyren}\ \emph {et~al.}(2007)\citenamefont {Reyren},
  \citenamefont {Thiel}, \citenamefont {Caviglia}, \citenamefont {Kourkoutis},
  \citenamefont {Hammerl}, \citenamefont {Richter}, \citenamefont {Schneider},
  \citenamefont {Kopp}, \citenamefont {Ruetschi}, \citenamefont {Jaccard},
  \citenamefont {Gabay}, \citenamefont {Muller}, \citenamefont {Triscone},\
  and\ \citenamefont {Mannhart}}]{reyren2007}%
  \BibitemOpen
  \bibfield  {author} {\bibinfo {author} {\bibfnamefont {N.}~\bibnamefont
  {Reyren}}, \bibinfo {author} {\bibfnamefont {S.}~\bibnamefont {Thiel}},
  \bibinfo {author} {\bibfnamefont {A.~D.}\ \bibnamefont {Caviglia}}, \bibinfo
  {author} {\bibfnamefont {L.~F.}\ \bibnamefont {Kourkoutis}}, \bibinfo
  {author} {\bibfnamefont {G.}~\bibnamefont {Hammerl}}, \bibinfo {author}
  {\bibfnamefont {C.}~\bibnamefont {Richter}}, \bibinfo {author} {\bibfnamefont
  {C.~W.}\ \bibnamefont {Schneider}}, \bibinfo {author} {\bibfnamefont
  {T.}~\bibnamefont {Kopp}}, \bibinfo {author} {\bibfnamefont {A.-S.}\
  \bibnamefont {Ruetschi}}, \bibinfo {author} {\bibfnamefont {D.}~\bibnamefont
  {Jaccard}}, \bibinfo {author} {\bibfnamefont {M.}~\bibnamefont {Gabay}},
  \bibinfo {author} {\bibfnamefont {D.~A.}\ \bibnamefont {Muller}}, \bibinfo
  {author} {\bibfnamefont {J.-M.}\ \bibnamefont {Triscone}},\ and\ \bibinfo
  {author} {\bibfnamefont {J.}~\bibnamefont {Mannhart}},\ }\href@noop {}
  {\bibfield  {journal} {\bibinfo  {journal} {Science (New York, N.Y.)}\
  }\textbf {\bibinfo {volume} {317}},\ \bibinfo {pages} {1196} (\bibinfo {year}
  {2007})}\BibitemShut {NoStop}%
\bibitem [{\citenamefont {Brinkman}\ \emph {et~al.}(2007)\citenamefont
  {Brinkman}, \citenamefont {Huijben}, \citenamefont {Van~Zalk}, \citenamefont
  {Huijben}, \citenamefont {Zeitler}, \citenamefont {Maan}, \citenamefont {Van
  Der~Wiel}, \citenamefont {Rijnders}, \citenamefont {Blank},\ and\
  \citenamefont {Hilgenkamp}}]{Brinkman2007}%
  \BibitemOpen
  \bibfield  {author} {\bibinfo {author} {\bibfnamefont {A.}~\bibnamefont
  {Brinkman}}, \bibinfo {author} {\bibfnamefont {M.}~\bibnamefont {Huijben}},
  \bibinfo {author} {\bibfnamefont {M.}~\bibnamefont {Van~Zalk}}, \bibinfo
  {author} {\bibfnamefont {J.}~\bibnamefont {Huijben}}, \bibinfo {author}
  {\bibfnamefont {U.}~\bibnamefont {Zeitler}}, \bibinfo {author} {\bibfnamefont
  {J.~C.}\ \bibnamefont {Maan}}, \bibinfo {author} {\bibfnamefont {W.~G.}\
  \bibnamefont {Van Der~Wiel}}, \bibinfo {author} {\bibfnamefont
  {G.}~\bibnamefont {Rijnders}}, \bibinfo {author} {\bibfnamefont {D.~H.}\
  \bibnamefont {Blank}},\ and\ \bibinfo {author} {\bibfnamefont
  {H.}~\bibnamefont {Hilgenkamp}},\ }\href@noop {} {\bibfield  {journal}
  {\bibinfo  {journal} {Nat. Mater.}\ }\textbf {\bibinfo {volume} {6}},\
  \bibinfo {pages} {493} (\bibinfo {year} {2007})}\BibitemShut {NoStop}%
\bibitem [{\citenamefont {Li}\ \emph {et~al.}(2011)\citenamefont {Li},
  \citenamefont {Richter}, \citenamefont {Mannhart},\ and\ \citenamefont
  {Ashoori}}]{Li2011}%
  \BibitemOpen
  \bibfield  {author} {\bibinfo {author} {\bibfnamefont {L.}~\bibnamefont
  {Li}}, \bibinfo {author} {\bibfnamefont {C.}~\bibnamefont {Richter}},
  \bibinfo {author} {\bibfnamefont {J.}~\bibnamefont {Mannhart}},\ and\
  \bibinfo {author} {\bibfnamefont {R.~C.}\ \bibnamefont {Ashoori}},\
  }\href@noop {} {\bibfield  {journal} {\bibinfo  {journal} {Nat. Phys.}\
  }\textbf {\bibinfo {volume} {7}},\ \bibinfo {pages} {762} (\bibinfo {year}
  {2011})}\BibitemShut {NoStop}%
\bibitem [{\citenamefont {Caviglia}\ \emph {et~al.}(2008)\citenamefont
  {Caviglia}, \citenamefont {Gariglio}, \citenamefont {Reyren}, \citenamefont
  {Jaccard}, \citenamefont {Schneider}, \citenamefont {Gabay}, \citenamefont
  {Thiel}, \citenamefont {Hammerl}, \citenamefont {Mannhart},\ and\
  \citenamefont {Triscone}}]{Caviglia2008}%
  \BibitemOpen
  \bibfield  {author} {\bibinfo {author} {\bibfnamefont {A.~D.}\ \bibnamefont
  {Caviglia}}, \bibinfo {author} {\bibfnamefont {S.}~\bibnamefont {Gariglio}},
  \bibinfo {author} {\bibfnamefont {N.}~\bibnamefont {Reyren}}, \bibinfo
  {author} {\bibfnamefont {D.}~\bibnamefont {Jaccard}}, \bibinfo {author}
  {\bibfnamefont {T.}~\bibnamefont {Schneider}}, \bibinfo {author}
  {\bibfnamefont {M.}~\bibnamefont {Gabay}}, \bibinfo {author} {\bibfnamefont
  {S.}~\bibnamefont {Thiel}}, \bibinfo {author} {\bibfnamefont
  {G.}~\bibnamefont {Hammerl}}, \bibinfo {author} {\bibfnamefont
  {J.}~\bibnamefont {Mannhart}},\ and\ \bibinfo {author} {\bibfnamefont
  {J.-M.}\ \bibnamefont {Triscone}},\ }\href@noop {} {\bibfield  {journal}
  {\bibinfo  {journal} {Nature}\ }\textbf {\bibinfo {volume} {456}},\ \bibinfo
  {pages} {624} (\bibinfo {year} {2008})}\BibitemShut {NoStop}%
\bibitem [{\citenamefont {Caviglia}\ \emph
  {et~al.}(2010{\natexlab{a}})\citenamefont {Caviglia}, \citenamefont
  {Gariglio}, \citenamefont {Cancellieri}, \citenamefont {Sacépé},
  \citenamefont {Fête}, \citenamefont {Reyren}, \citenamefont {Gabay},
  \citenamefont {Morpurgo},\ and\ \citenamefont {Triscone}}]{Caviglia2010}%
  \BibitemOpen
  \bibfield  {author} {\bibinfo {author} {\bibfnamefont {A.~D.}\ \bibnamefont
  {Caviglia}}, \bibinfo {author} {\bibfnamefont {S.}~\bibnamefont {Gariglio}},
  \bibinfo {author} {\bibfnamefont {C.}~\bibnamefont {Cancellieri}}, \bibinfo
  {author} {\bibfnamefont {B.}~\bibnamefont {Sacépé}}, \bibinfo {author}
  {\bibfnamefont {A.}~\bibnamefont {Fête}}, \bibinfo {author} {\bibfnamefont
  {N.}~\bibnamefont {Reyren}}, \bibinfo {author} {\bibfnamefont
  {M.}~\bibnamefont {Gabay}}, \bibinfo {author} {\bibfnamefont {A.~F.}\
  \bibnamefont {Morpurgo}},\ and\ \bibinfo {author} {\bibfnamefont {J.~M.}\
  \bibnamefont {Triscone}},\ }\href@noop {} {\bibfield  {journal} {\bibinfo
  {journal} {Phys. Rev. Lett.}\ }\textbf {\bibinfo {volume} {105}},\ \bibinfo
  {pages} {236802} (\bibinfo {year} {2010}{\natexlab{a}})}\BibitemShut
  {NoStop}%
\bibitem [{\citenamefont {Caviglia}\ \emph
  {et~al.}(2010{\natexlab{b}})\citenamefont {Caviglia}, \citenamefont {Gabay},
  \citenamefont {Gariglio}, \citenamefont {Reyren}, \citenamefont
  {Cancellieri},\ and\ \citenamefont {Triscone}}]{Caviglia2010a}%
  \BibitemOpen
  \bibfield  {author} {\bibinfo {author} {\bibfnamefont {A.~D.}\ \bibnamefont
  {Caviglia}}, \bibinfo {author} {\bibfnamefont {M.}~\bibnamefont {Gabay}},
  \bibinfo {author} {\bibfnamefont {S.}~\bibnamefont {Gariglio}}, \bibinfo
  {author} {\bibfnamefont {N.}~\bibnamefont {Reyren}}, \bibinfo {author}
  {\bibfnamefont {C.}~\bibnamefont {Cancellieri}},\ and\ \bibinfo {author}
  {\bibfnamefont {J.-M.}\ \bibnamefont {Triscone}},\ }\href@noop {} {\bibfield
  {journal} {\bibinfo  {journal} {Phys. Rev. Lett.}\ }\textbf {\bibinfo
  {volume} {104}},\ \bibinfo {pages} {126803} (\bibinfo {year}
  {2010}{\natexlab{b}})}\BibitemShut {NoStop}%
\bibitem [{\citenamefont {Esswein}\ \emph {et~al.}(2022)\citenamefont
  {Esswein}, \citenamefont {Spaldin},\ and\ \citenamefont
  {Ferroelectric}}]{Esswein2022}%
  \BibitemOpen
  \bibfield  {author} {\bibinfo {author} {\bibfnamefont {T.}~\bibnamefont
  {Esswein}}, \bibinfo {author} {\bibfnamefont {N.~A.}\ \bibnamefont
  {Spaldin}},\ and\ \bibinfo {author} {\bibfnamefont {q.~p.}\ \bibnamefont
  {Ferroelectric}},\ }\href@noop {} {\bibfield  {journal} {\bibinfo  {journal}
  {Phys. Rev. Res.}\ }\textbf {\bibinfo {volume} {4}},\ \bibinfo {pages}
  {033020} (\bibinfo {year} {2022})}\BibitemShut {NoStop}%
\bibitem [{\citenamefont {Fujishita}\ \emph {et~al.}(2016)\citenamefont
  {Fujishita}, \citenamefont {Kitazawa}, \citenamefont {Saito}, \citenamefont
  {Ishisaka}, \citenamefont {Okamoto},\ and\ \citenamefont
  {Yamaguchi}}]{Fujishita2016}%
  \BibitemOpen
  \bibfield  {author} {\bibinfo {author} {\bibfnamefont {H.}~\bibnamefont
  {Fujishita}}, \bibinfo {author} {\bibfnamefont {S.}~\bibnamefont {Kitazawa}},
  \bibinfo {author} {\bibfnamefont {M.}~\bibnamefont {Saito}}, \bibinfo
  {author} {\bibfnamefont {R.}~\bibnamefont {Ishisaka}}, \bibinfo {author}
  {\bibfnamefont {H.}~\bibnamefont {Okamoto}},\ and\ \bibinfo {author}
  {\bibfnamefont {T.}~\bibnamefont {Yamaguchi}},\ }\href@noop {} {\bibfield
  {journal} {\bibinfo  {journal} {J. Phys. Soc. Jpn.}\ }\textbf {\bibinfo
  {volume} {85}},\ \bibinfo {pages} {074703} (\bibinfo {year}
  {2016})}\BibitemShut {NoStop}%
\bibitem [{\citenamefont {Liu}\ \emph {et~al.}(2021)\citenamefont {Liu},
  \citenamefont {Yan}, \citenamefont {Jin}, \citenamefont {Ma}, \citenamefont
  {Hsiao}, \citenamefont {Lin}, \citenamefont {Bretz-Sullivan}, \citenamefont
  {Zhou}, \citenamefont {Pearson}, \citenamefont {Fisher}, \citenamefont
  {Jiang}, \citenamefont {Han}, \citenamefont {Zuo}, \citenamefont {Wen},
  \citenamefont {Fong}, \citenamefont {Sun}, \citenamefont {Zhou},\ and\
  \citenamefont {Bhattacharya}}]{Liu2021}%
  \BibitemOpen
  \bibfield  {author} {\bibinfo {author} {\bibfnamefont {C.}~\bibnamefont
  {Liu}}, \bibinfo {author} {\bibfnamefont {X.}~\bibnamefont {Yan}}, \bibinfo
  {author} {\bibfnamefont {D.}~\bibnamefont {Jin}}, \bibinfo {author}
  {\bibfnamefont {Y.}~\bibnamefont {Ma}}, \bibinfo {author} {\bibfnamefont
  {H.-W.}\ \bibnamefont {Hsiao}}, \bibinfo {author} {\bibfnamefont
  {Y.}~\bibnamefont {Lin}}, \bibinfo {author} {\bibfnamefont {T.~M.}\
  \bibnamefont {Bretz-Sullivan}}, \bibinfo {author} {\bibfnamefont
  {X.}~\bibnamefont {Zhou}}, \bibinfo {author} {\bibfnamefont {J.}~\bibnamefont
  {Pearson}}, \bibinfo {author} {\bibfnamefont {B.}~\bibnamefont {Fisher}},
  \bibinfo {author} {\bibfnamefont {J.~S.}\ \bibnamefont {Jiang}}, \bibinfo
  {author} {\bibfnamefont {W.}~\bibnamefont {Han}}, \bibinfo {author}
  {\bibfnamefont {J.-M.}\ \bibnamefont {Zuo}}, \bibinfo {author} {\bibfnamefont
  {J.}~\bibnamefont {Wen}}, \bibinfo {author} {\bibfnamefont {D.~D.}\
  \bibnamefont {Fong}}, \bibinfo {author} {\bibfnamefont {J.}~\bibnamefont
  {Sun}}, \bibinfo {author} {\bibfnamefont {H.}~\bibnamefont {Zhou}},\ and\
  \bibinfo {author} {\bibfnamefont {A.}~\bibnamefont {Bhattacharya}},\
  }\href@noop {} {\bibfield  {journal} {\bibinfo  {journal} {Science}\ }\textbf
  {\bibinfo {volume} {371}},\ \bibinfo {pages} {716} (\bibinfo {year}
  {2021})}\BibitemShut {NoStop}%
\bibitem [{\citenamefont {Chen}\ \emph
  {et~al.}(2021{\natexlab{a}})\citenamefont {Chen}, \citenamefont {Liu},
  \citenamefont {Zhang}, \citenamefont {Liu}, \citenamefont {Tian},
  \citenamefont {Sun}, \citenamefont {Zhang}, \citenamefont {Zhou},
  \citenamefont {Sun},\ and\ \citenamefont {Xie}}]{Chen2021a}%
  \BibitemOpen
  \bibfield  {author} {\bibinfo {author} {\bibfnamefont {Z.}~\bibnamefont
  {Chen}}, \bibinfo {author} {\bibfnamefont {Y.}~\bibnamefont {Liu}}, \bibinfo
  {author} {\bibfnamefont {H.}~\bibnamefont {Zhang}}, \bibinfo {author}
  {\bibfnamefont {Z.}~\bibnamefont {Liu}}, \bibinfo {author} {\bibfnamefont
  {H.}~\bibnamefont {Tian}}, \bibinfo {author} {\bibfnamefont {Y.}~\bibnamefont
  {Sun}}, \bibinfo {author} {\bibfnamefont {M.}~\bibnamefont {Zhang}}, \bibinfo
  {author} {\bibfnamefont {Y.}~\bibnamefont {Zhou}}, \bibinfo {author}
  {\bibfnamefont {J.}~\bibnamefont {Sun}},\ and\ \bibinfo {author}
  {\bibfnamefont {Y.}~\bibnamefont {Xie}},\ }\href@noop {} {\bibfield
  {journal} {\bibinfo  {journal} {Science}\ }\textbf {\bibinfo {volume}
  {372}},\ \bibinfo {pages} {721} (\bibinfo {year}
  {2021}{\natexlab{a}})}\BibitemShut {NoStop}%
\bibitem [{\citenamefont {Liu}\ \emph {et~al.}(2023)\citenamefont {Liu},
  \citenamefont {Zhou}, \citenamefont {Hong}, \citenamefont {Fisher},
  \citenamefont {Zheng}, \citenamefont {Pearson}, \citenamefont {Jiang},
  \citenamefont {Jin}, \citenamefont {Norman},\ and\ \citenamefont
  {Bhattacharya}}]{Liu2023}%
  \BibitemOpen
  \bibfield  {author} {\bibinfo {author} {\bibfnamefont {C.}~\bibnamefont
  {Liu}}, \bibinfo {author} {\bibfnamefont {X.}~\bibnamefont {Zhou}}, \bibinfo
  {author} {\bibfnamefont {D.}~\bibnamefont {Hong}}, \bibinfo {author}
  {\bibfnamefont {B.}~\bibnamefont {Fisher}}, \bibinfo {author} {\bibfnamefont
  {H.}~\bibnamefont {Zheng}}, \bibinfo {author} {\bibfnamefont
  {J.}~\bibnamefont {Pearson}}, \bibinfo {author} {\bibfnamefont {J.~S.}\
  \bibnamefont {Jiang}}, \bibinfo {author} {\bibfnamefont {D.}~\bibnamefont
  {Jin}}, \bibinfo {author} {\bibfnamefont {M.~R.}\ \bibnamefont {Norman}},\
  and\ \bibinfo {author} {\bibfnamefont {A.}~\bibnamefont {Bhattacharya}},\
  }\href@noop {} {\bibfield  {journal} {\bibinfo  {journal} {Nat. Commun.}\
  }\textbf {\bibinfo {volume} {14}},\ \bibinfo {pages} {951} (\bibinfo {year}
  {2023})}\BibitemShut {NoStop}%
\bibitem [{\citenamefont {Chen}\ \emph
  {et~al.}(2021{\natexlab{b}})\citenamefont {Chen}, \citenamefont {Liu},
  \citenamefont {Sun}, \citenamefont {Chen}, \citenamefont {Liu}, \citenamefont
  {Zhang}, \citenamefont {Li}, \citenamefont {Zhang}, \citenamefont {Hong},
  \citenamefont {Ren}, \citenamefont {Zhang}, \citenamefont {Tian},
  \citenamefont {Zhou}, \citenamefont {Sun},\ and\ \citenamefont
  {Xie}}]{Chen2021b}%
  \BibitemOpen
  \bibfield  {author} {\bibinfo {author} {\bibfnamefont {Z.}~\bibnamefont
  {Chen}}, \bibinfo {author} {\bibfnamefont {Z.}~\bibnamefont {Liu}}, \bibinfo
  {author} {\bibfnamefont {Y.}~\bibnamefont {Sun}}, \bibinfo {author}
  {\bibfnamefont {X.}~\bibnamefont {Chen}}, \bibinfo {author} {\bibfnamefont
  {Y.}~\bibnamefont {Liu}}, \bibinfo {author} {\bibfnamefont {H.}~\bibnamefont
  {Zhang}}, \bibinfo {author} {\bibfnamefont {H.}~\bibnamefont {Li}}, \bibinfo
  {author} {\bibfnamefont {M.}~\bibnamefont {Zhang}}, \bibinfo {author}
  {\bibfnamefont {S.}~\bibnamefont {Hong}}, \bibinfo {author} {\bibfnamefont
  {T.}~\bibnamefont {Ren}}, \bibinfo {author} {\bibfnamefont {C.}~\bibnamefont
  {Zhang}}, \bibinfo {author} {\bibfnamefont {H.}~\bibnamefont {Tian}},
  \bibinfo {author} {\bibfnamefont {Y.}~\bibnamefont {Zhou}}, \bibinfo {author}
  {\bibfnamefont {J.}~\bibnamefont {Sun}},\ and\ \bibinfo {author}
  {\bibfnamefont {Y.}~\bibnamefont {Xie}},\ }\href@noop {} {\bibfield
  {journal} {\bibinfo  {journal} {Phys. Rev. Lett.}\ }\textbf {\bibinfo
  {volume} {126}},\ \bibinfo {pages} {026802} (\bibinfo {year}
  {2021}{\natexlab{b}})}\BibitemShut {NoStop}%
\bibitem [{\citenamefont {Hua}\ \emph {et~al.}(2024)\citenamefont {Hua},
  \citenamefont {Zeng}, \citenamefont {Meng}, \citenamefont {Yao},
  \citenamefont {Huang}, \citenamefont {Long}, \citenamefont {Li},
  \citenamefont {Wang}, \citenamefont {Wang}, \citenamefont {Wu}, \citenamefont
  {Weng}, \citenamefont {Wang}, \citenamefont {Liu}, \citenamefont {Xiang},\
  and\ \citenamefont {Chen}}]{Hua2024}%
  \BibitemOpen
  \bibfield  {author} {\bibinfo {author} {\bibfnamefont {X.}~\bibnamefont
  {Hua}}, \bibinfo {author} {\bibfnamefont {Z.}~\bibnamefont {Zeng}}, \bibinfo
  {author} {\bibfnamefont {F.}~\bibnamefont {Meng}}, \bibinfo {author}
  {\bibfnamefont {H.}~\bibnamefont {Yao}}, \bibinfo {author} {\bibfnamefont
  {Z.}~\bibnamefont {Huang}}, \bibinfo {author} {\bibfnamefont
  {X.}~\bibnamefont {Long}}, \bibinfo {author} {\bibfnamefont {Z.}~\bibnamefont
  {Li}}, \bibinfo {author} {\bibfnamefont {Y.}~\bibnamefont {Wang}}, \bibinfo
  {author} {\bibfnamefont {Z.}~\bibnamefont {Wang}}, \bibinfo {author}
  {\bibfnamefont {T.}~\bibnamefont {Wu}}, \bibinfo {author} {\bibfnamefont
  {Z.}~\bibnamefont {Weng}}, \bibinfo {author} {\bibfnamefont {Y.}~\bibnamefont
  {Wang}}, \bibinfo {author} {\bibfnamefont {Z.}~\bibnamefont {Liu}}, \bibinfo
  {author} {\bibfnamefont {Z.}~\bibnamefont {Xiang}},\ and\ \bibinfo {author}
  {\bibfnamefont {X.}~\bibnamefont {Chen}},\ }\href@noop {} {\bibfield
  {journal} {\bibinfo  {journal} {Nat. Phys.}\ }\textbf {\bibinfo {volume}
  {20}},\ \bibinfo {pages} {957} (\bibinfo {year} {2024})}\BibitemShut
  {NoStop}%
\bibitem [{\citenamefont {Arnault}\ \emph {et~al.}(2023)\citenamefont
  {Arnault}, \citenamefont {Al-Tawhid}, \citenamefont {Salmani-Rezaie},
  \citenamefont {Muller}, \citenamefont {Kumah}, \citenamefont {Bahramy},
  \citenamefont {Finkelstein},\ and\ \citenamefont {Ahadi}}]{Arnault2023a}%
  \BibitemOpen
  \bibfield  {author} {\bibinfo {author} {\bibfnamefont {E.~G.}\ \bibnamefont
  {Arnault}}, \bibinfo {author} {\bibfnamefont {A.~H.}\ \bibnamefont
  {Al-Tawhid}}, \bibinfo {author} {\bibfnamefont {S.}~\bibnamefont
  {Salmani-Rezaie}}, \bibinfo {author} {\bibfnamefont {D.~A.}\ \bibnamefont
  {Muller}}, \bibinfo {author} {\bibfnamefont {D.~P.}\ \bibnamefont {Kumah}},
  \bibinfo {author} {\bibfnamefont {M.~S.}\ \bibnamefont {Bahramy}}, \bibinfo
  {author} {\bibfnamefont {G.}~\bibnamefont {Finkelstein}},\ and\ \bibinfo
  {author} {\bibfnamefont {K.}~\bibnamefont {Ahadi}},\ }\href@noop {}
  {\bibfield  {journal} {\bibinfo  {journal} {Sci. Adv.}\ }\textbf {\bibinfo
  {volume} {9}},\ \bibinfo {pages} {eadf1414} (\bibinfo {year}
  {2023})}\BibitemShut {NoStop}%
\bibitem [{\citenamefont {Bruno}\ \emph {et~al.}(2019)\citenamefont {Bruno},
  \citenamefont {McKeown~Walker}, \citenamefont {Ricc{\`o}}, \citenamefont
  {De~La~Torre}, \citenamefont {Wang}, \citenamefont {Tamai}, \citenamefont
  {Kim}, \citenamefont {Hoesch}, \citenamefont {Bahramy},\ and\ \citenamefont
  {Baumberger}}]{Bruno2019a}%
  \BibitemOpen
  \bibfield  {author} {\bibinfo {author} {\bibfnamefont {F.~Y.}\ \bibnamefont
  {Bruno}}, \bibinfo {author} {\bibfnamefont {S.}~\bibnamefont
  {McKeown~Walker}}, \bibinfo {author} {\bibfnamefont {S.}~\bibnamefont
  {Ricc{\`o}}}, \bibinfo {author} {\bibfnamefont {A.}~\bibnamefont
  {De~La~Torre}}, \bibinfo {author} {\bibfnamefont {Z.}~\bibnamefont {Wang}},
  \bibinfo {author} {\bibfnamefont {A.}~\bibnamefont {Tamai}}, \bibinfo
  {author} {\bibfnamefont {T.~K.}\ \bibnamefont {Kim}}, \bibinfo {author}
  {\bibfnamefont {M.}~\bibnamefont {Hoesch}}, \bibinfo {author} {\bibfnamefont
  {M.~S.}\ \bibnamefont {Bahramy}},\ and\ \bibinfo {author} {\bibfnamefont
  {F.}~\bibnamefont {Baumberger}},\ }\href@noop {} {\bibfield  {journal}
  {\bibinfo  {journal} {Adv. Electron. Mater.}\ }\textbf {\bibinfo {volume}
  {5}},\ \bibinfo {pages} {1800860} (\bibinfo {year} {2019})}\BibitemShut
  {NoStop}%
\bibitem [{\citenamefont {Zhai}\ \emph {et~al.}(2025)\citenamefont {Zhai},
  \citenamefont {Oh}, \citenamefont {Liu}, \citenamefont {Pan}, \citenamefont
  {Nagaosa}, \citenamefont {He},\ and\ \citenamefont {Shen}}]{Zhai2025}%
  \BibitemOpen
  \bibfield  {author} {\bibinfo {author} {\bibfnamefont {J.}~\bibnamefont
  {Zhai}}, \bibinfo {author} {\bibfnamefont {T.}~\bibnamefont {Oh}}, \bibinfo
  {author} {\bibfnamefont {H.}~\bibnamefont {Liu}}, \bibinfo {author}
  {\bibfnamefont {C.}~\bibnamefont {Pan}}, \bibinfo {author} {\bibfnamefont
  {N.}~\bibnamefont {Nagaosa}}, \bibinfo {author} {\bibfnamefont
  {P.}~\bibnamefont {He}},\ and\ \bibinfo {author} {\bibfnamefont
  {J.}~\bibnamefont {Shen}},\ }\href@noop {} {\bibfield  {journal} {\bibinfo
  {journal} {Phys. Rev. Lett.}\ }\textbf {\bibinfo {volume} {134}},\ \bibinfo
  {pages} {236303} (\bibinfo {year} {2025})}\BibitemShut {NoStop}%
\bibitem [{\citenamefont {Zhang}\ \emph {et~al.}(2023)\citenamefont {Zhang},
  \citenamefont {Wang}, \citenamefont {Wang}, \citenamefont {Li}, \citenamefont
  {Huang}, \citenamefont {Yang}, \citenamefont {Xue}, \citenamefont {Ning},
  \citenamefont {Wu}, \citenamefont {Xu}, \citenamefont {Song}, \citenamefont
  {An}, \citenamefont {Zheng}, \citenamefont {Shen}, \citenamefont {Li},
  \citenamefont {Chen},\ and\ \citenamefont {Li}}]{Zhang2023}%
  \BibitemOpen
  \bibfield  {author} {\bibinfo {author} {\bibfnamefont {G.}~\bibnamefont
  {Zhang}}, \bibinfo {author} {\bibfnamefont {L.}~\bibnamefont {Wang}},
  \bibinfo {author} {\bibfnamefont {J.}~\bibnamefont {Wang}}, \bibinfo {author}
  {\bibfnamefont {G.}~\bibnamefont {Li}}, \bibinfo {author} {\bibfnamefont
  {G.}~\bibnamefont {Huang}}, \bibinfo {author} {\bibfnamefont
  {G.}~\bibnamefont {Yang}}, \bibinfo {author} {\bibfnamefont {H.}~\bibnamefont
  {Xue}}, \bibinfo {author} {\bibfnamefont {Z.}~\bibnamefont {Ning}}, \bibinfo
  {author} {\bibfnamefont {Y.}~\bibnamefont {Wu}}, \bibinfo {author}
  {\bibfnamefont {J.-P.}\ \bibnamefont {Xu}}, \bibinfo {author} {\bibfnamefont
  {Y.}~\bibnamefont {Song}}, \bibinfo {author} {\bibfnamefont {Z.}~\bibnamefont
  {An}}, \bibinfo {author} {\bibfnamefont {C.}~\bibnamefont {Zheng}}, \bibinfo
  {author} {\bibfnamefont {J.}~\bibnamefont {Shen}}, \bibinfo {author}
  {\bibfnamefont {J.}~\bibnamefont {Li}}, \bibinfo {author} {\bibfnamefont
  {Y.}~\bibnamefont {Chen}},\ and\ \bibinfo {author} {\bibfnamefont
  {W.}~\bibnamefont {Li}},\ }\href@noop {} {\bibfield  {journal} {\bibinfo
  {journal} {Nat. Commun.}\ }\textbf {\bibinfo {volume} {14}},\ \bibinfo
  {pages} {3046} (\bibinfo {year} {2023})}\BibitemShut {NoStop}%
\bibitem [{Sup()}]{Suppl}%
  \BibitemOpen
  \href@noop {} {}\bibinfo {note} {See Supplemental Material at [] for sample
  preparation, electric transport measurements, definition of $r$, simulation
  of vortex motion in superconducting stripes and comparison of {S1} and {S2}.
  The Supplemental Material also contains Refs. [36, 37].}\BibitemShut {Stop}%
\bibitem [{\citenamefont {Pang}\ \emph {et~al.}()\citenamefont {Pang},
  \citenamefont {Davidson}, \citenamefont {Li}, \citenamefont {Oudah},
  \citenamefont {Moen}, \citenamefont {Smit}, \citenamefont {Suen},
  \citenamefont {Godin}, \citenamefont {Gorovikov}, \citenamefont {Zonno},
  \citenamefont {Zhdanovich}, \citenamefont {Levy}, \citenamefont {Michiardi},
  \citenamefont {Hallas}, \citenamefont {Damascelli},\ and\ \citenamefont
  {Zou}}]{Pang2025}%
  \BibitemOpen
  \bibfield  {author} {\bibinfo {author} {\bibfnamefont {C.~S.~B.}\
  \bibnamefont {Pang}}, \bibinfo {author} {\bibfnamefont {B.~A.}\ \bibnamefont
  {Davidson}}, \bibinfo {author} {\bibfnamefont {F.}~\bibnamefont {Li}},
  \bibinfo {author} {\bibfnamefont {M.}~\bibnamefont {Oudah}}, \bibinfo
  {author} {\bibfnamefont {P.}~\bibnamefont {Moen}}, \bibinfo {author}
  {\bibfnamefont {S.}~\bibnamefont {Smit}}, \bibinfo {author} {\bibfnamefont
  {C.~T.}\ \bibnamefont {Suen}}, \bibinfo {author} {\bibfnamefont
  {S.}~\bibnamefont {Godin}}, \bibinfo {author} {\bibfnamefont {S.~A.}\
  \bibnamefont {Gorovikov}}, \bibinfo {author} {\bibfnamefont {M.}~\bibnamefont
  {Zonno}}, \bibinfo {author} {\bibfnamefont {S.}~\bibnamefont {Zhdanovich}},
  \bibinfo {author} {\bibfnamefont {G.}~\bibnamefont {Levy}}, \bibinfo {author}
  {\bibfnamefont {M.}~\bibnamefont {Michiardi}}, \bibinfo {author}
  {\bibfnamefont {A.~M.}\ \bibnamefont {Hallas}}, \bibinfo {author}
  {\bibfnamefont {A.}~\bibnamefont {Damascelli}},\ and\ \bibinfo {author}
  {\bibfnamefont {K.}~\bibnamefont {Zou}},\ }\bibinfo {note} {in
  preparation}\BibitemShut {NoStop}%
\bibitem [{\citenamefont {Tinkham}(1963)}]{Tinkham1963}%
  \BibitemOpen
  \bibfield  {author} {\bibinfo {author} {\bibfnamefont {M.}~\bibnamefont
  {Tinkham}},\ }\href@noop {} {\bibfield  {journal} {\bibinfo  {journal} {Phys.
  Rev.}\ }\textbf {\bibinfo {volume} {129}},\ \bibinfo {pages} {2413} (\bibinfo
  {year} {1963})}\BibitemShut {NoStop}%
\bibitem [{\citenamefont {Jiang}\ \emph {et~al.}(2020)\citenamefont {Jiang},
  \citenamefont {Yuan}, \citenamefont {Wu}, \citenamefont {Wei}, \citenamefont
  {Mu}, \citenamefont {An},\ and\ \citenamefont {Li}}]{Jiang2020}%
  \BibitemOpen
  \bibfield  {author} {\bibinfo {author} {\bibfnamefont {D.}~\bibnamefont
  {Jiang}}, \bibinfo {author} {\bibfnamefont {T.}~\bibnamefont {Yuan}},
  \bibinfo {author} {\bibfnamefont {Y.}~\bibnamefont {Wu}}, \bibinfo {author}
  {\bibfnamefont {X.}~\bibnamefont {Wei}}, \bibinfo {author} {\bibfnamefont
  {G.}~\bibnamefont {Mu}}, \bibinfo {author} {\bibfnamefont {Z.}~\bibnamefont
  {An}},\ and\ \bibinfo {author} {\bibfnamefont {W.}~\bibnamefont {Li}},\
  }\href@noop {} {\bibfield  {journal} {\bibinfo  {journal} {ACS Appl. Mater.
  Interfaces}\ }\textbf {\bibinfo {volume} {12}},\ \bibinfo {pages} {49252}
  (\bibinfo {year} {2020})}\BibitemShut {NoStop}%
\bibitem [{\citenamefont {Zhang}\ \emph
  {et~al.}(2021{\natexlab{a}})\citenamefont {Zhang}, \citenamefont {Xing},
  \citenamefont {Meng}, \citenamefont {Yi}, \citenamefont {Gouchi},
  \citenamefont {Bhoi}, \citenamefont {Feng}, \citenamefont {Fan},
  \citenamefont {Zhou}, \citenamefont {Zhou}, \citenamefont {Uwatoko},\ and\
  \citenamefont {Shi}}]{Zhang2021a}%
  \BibitemOpen
  \bibfield  {author} {\bibinfo {author} {\bibfnamefont {Y.}~\bibnamefont
  {Zhang}}, \bibinfo {author} {\bibfnamefont {X.}~\bibnamefont {Xing}},
  \bibinfo {author} {\bibfnamefont {Y.}~\bibnamefont {Meng}}, \bibinfo {author}
  {\bibfnamefont {X.}~\bibnamefont {Yi}}, \bibinfo {author} {\bibfnamefont
  {J.}~\bibnamefont {Gouchi}}, \bibinfo {author} {\bibfnamefont
  {D.}~\bibnamefont {Bhoi}}, \bibinfo {author} {\bibfnamefont {J.}~\bibnamefont
  {Feng}}, \bibinfo {author} {\bibfnamefont {Y.}~\bibnamefont {Fan}}, \bibinfo
  {author} {\bibfnamefont {W.}~\bibnamefont {Zhou}}, \bibinfo {author}
  {\bibfnamefont {N.}~\bibnamefont {Zhou}}, \bibinfo {author} {\bibfnamefont
  {Y.}~\bibnamefont {Uwatoko}},\ and\ \bibinfo {author} {\bibfnamefont
  {Z.}~\bibnamefont {Shi}},\ }\href@noop {} {\bibfield  {journal} {\bibinfo
  {journal} {Chem. Mater.}\ }\textbf {\bibinfo {volume} {33}},\ \bibinfo
  {pages} {6752} (\bibinfo {year} {2021}{\natexlab{a}})}\BibitemShut {NoStop}%
\bibitem [{\citenamefont {Tinkham}(2015)}]{Tinkham2015a}%
  \BibitemOpen
  \bibfield  {author} {\bibinfo {author} {\bibfnamefont {M.}~\bibnamefont
  {Tinkham}},\ }\href@noop {} {\emph {\bibinfo {title} {Introduction to
  superconductivity}}},\ \bibinfo {edition} {2nd}\ ed.,\ Dover books on
  physics\ (\bibinfo  {publisher} {Dover Publ.},\ \bibinfo {address} {Mineola,
  NY},\ \bibinfo {year} {2015})\BibitemShut {NoStop}%
\bibitem [{\citenamefont {Zhang}\ \emph
  {et~al.}(2021{\natexlab{b}})\citenamefont {Zhang}, \citenamefont {Xu},
  \citenamefont {Huang}, \citenamefont {Zou}, \citenamefont {Ai}, \citenamefont
  {Liu}, \citenamefont {Leng}, \citenamefont {Jia}, \citenamefont {Zhang},
  \citenamefont {Zhao}, \citenamefont {Li}, \citenamefont {Yang}, \citenamefont
  {Liu}, \citenamefont {Haigh}, \citenamefont {Mao},\ and\ \citenamefont
  {Xiu}}]{Zhang2021}%
  \BibitemOpen
  \bibfield  {author} {\bibinfo {author} {\bibfnamefont {E.}~\bibnamefont
  {Zhang}}, \bibinfo {author} {\bibfnamefont {X.}~\bibnamefont {Xu}}, \bibinfo
  {author} {\bibfnamefont {C.}~\bibnamefont {Huang}}, \bibinfo {author}
  {\bibfnamefont {Y.-C.}\ \bibnamefont {Zou}}, \bibinfo {author} {\bibfnamefont
  {L.}~\bibnamefont {Ai}}, \bibinfo {author} {\bibfnamefont {S.}~\bibnamefont
  {Liu}}, \bibinfo {author} {\bibfnamefont {P.}~\bibnamefont {Leng}}, \bibinfo
  {author} {\bibfnamefont {Z.}~\bibnamefont {Jia}}, \bibinfo {author}
  {\bibfnamefont {Y.}~\bibnamefont {Zhang}}, \bibinfo {author} {\bibfnamefont
  {M.}~\bibnamefont {Zhao}}, \bibinfo {author} {\bibfnamefont {Z.}~\bibnamefont
  {Li}}, \bibinfo {author} {\bibfnamefont {Y.}~\bibnamefont {Yang}}, \bibinfo
  {author} {\bibfnamefont {J.}~\bibnamefont {Liu}}, \bibinfo {author}
  {\bibfnamefont {S.~J.}\ \bibnamefont {Haigh}}, \bibinfo {author}
  {\bibfnamefont {Z.}~\bibnamefont {Mao}},\ and\ \bibinfo {author}
  {\bibfnamefont {F.}~\bibnamefont {Xiu}},\ }\href@noop {} {\bibfield
  {journal} {\bibinfo  {journal} {Nano Lett.}\ }\textbf {\bibinfo {volume}
  {21}},\ \bibinfo {pages} {288} (\bibinfo {year}
  {2021}{\natexlab{b}})}\BibitemShut {NoStop}%
\bibitem [{\citenamefont {Xiong}\ \emph {et~al.}(1997)\citenamefont {Xiong},
  \citenamefont {Herzog},\ and\ \citenamefont {Dynes}}]{Xiong1997}%
  \BibitemOpen
  \bibfield  {author} {\bibinfo {author} {\bibfnamefont {P.}~\bibnamefont
  {Xiong}}, \bibinfo {author} {\bibfnamefont {A.~V.}\ \bibnamefont {Herzog}},\
  and\ \bibinfo {author} {\bibfnamefont {R.~C.}\ \bibnamefont {Dynes}},\
  }\href@noop {} {\bibfield  {journal} {\bibinfo  {journal} {Phys. Rev. Lett.}\
  }\textbf {\bibinfo {volume} {78}},\ \bibinfo {pages} {927} (\bibinfo {year}
  {1997})}\BibitemShut {NoStop}%
\bibitem [{\citenamefont {Berdiyorov}\ \emph {et~al.}(2012)\citenamefont
  {Berdiyorov}, \citenamefont {Milo{\v s}evi{\'c}}, \citenamefont {Latimer},
  \citenamefont {Xiao}, \citenamefont {Kwok},\ and\ \citenamefont
  {Peeters}}]{Berdiyorov2012}%
  \BibitemOpen
  \bibfield  {author} {\bibinfo {author} {\bibfnamefont {G.~R.}\ \bibnamefont
  {Berdiyorov}}, \bibinfo {author} {\bibfnamefont {M.~V.}\ \bibnamefont
  {Milo{\v s}evi{\'c}}}, \bibinfo {author} {\bibfnamefont {M.~L.}\ \bibnamefont
  {Latimer}}, \bibinfo {author} {\bibfnamefont {Z.~L.}\ \bibnamefont {Xiao}},
  \bibinfo {author} {\bibfnamefont {W.~K.}\ \bibnamefont {Kwok}},\ and\
  \bibinfo {author} {\bibfnamefont {F.~M.}\ \bibnamefont {Peeters}},\
  }\href@noop {} {\bibfield  {journal} {\bibinfo  {journal} {Phys. Rev. Lett.}\
  }\textbf {\bibinfo {volume} {109}},\ \bibinfo {pages} {057004} (\bibinfo
  {year} {2012})}\BibitemShut {NoStop}%
\bibitem [{\citenamefont {C{\'o}rdoba}\ \emph {et~al.}(2013)\citenamefont
  {C{\'o}rdoba}, \citenamefont {Baturina}, \citenamefont {Ses{\'e}},
  \citenamefont {Mironov}, \citenamefont {De~Teresa}, \citenamefont {Ibarra},
  \citenamefont {Nasimov}, \citenamefont {Gutakovskii}, \citenamefont
  {Latyshev}, \citenamefont {Guillam{\'o}n}, \citenamefont {Suderow},
  \citenamefont {Vieira}, \citenamefont {Baklanov}, \citenamefont {Palacios},\
  and\ \citenamefont {Vinokur}}]{Cordoba2013}%
  \BibitemOpen
  \bibfield  {author} {\bibinfo {author} {\bibfnamefont {R.}~\bibnamefont
  {C{\'o}rdoba}}, \bibinfo {author} {\bibfnamefont {T.~I.}\ \bibnamefont
  {Baturina}}, \bibinfo {author} {\bibfnamefont {J.}~\bibnamefont {Ses{\'e}}},
  \bibinfo {author} {\bibfnamefont {A.~Y.}\ \bibnamefont {Mironov}}, \bibinfo
  {author} {\bibfnamefont {J.~M.}\ \bibnamefont {De~Teresa}}, \bibinfo {author}
  {\bibfnamefont {M.~R.}\ \bibnamefont {Ibarra}}, \bibinfo {author}
  {\bibfnamefont {D.~A.}\ \bibnamefont {Nasimov}}, \bibinfo {author}
  {\bibfnamefont {A.~K.}\ \bibnamefont {Gutakovskii}}, \bibinfo {author}
  {\bibfnamefont {A.~V.}\ \bibnamefont {Latyshev}}, \bibinfo {author}
  {\bibfnamefont {I.}~\bibnamefont {Guillam{\'o}n}}, \bibinfo {author}
  {\bibfnamefont {H.}~\bibnamefont {Suderow}}, \bibinfo {author} {\bibfnamefont
  {S.}~\bibnamefont {Vieira}}, \bibinfo {author} {\bibfnamefont {M.~R.}\
  \bibnamefont {Baklanov}}, \bibinfo {author} {\bibfnamefont {J.~J.}\
  \bibnamefont {Palacios}},\ and\ \bibinfo {author} {\bibfnamefont {V.~M.}\
  \bibnamefont {Vinokur}},\ }\href@noop {} {\bibfield  {journal} {\bibinfo
  {journal} {Nat. Commun.}\ }\textbf {\bibinfo {volume} {4}},\ \bibinfo {pages}
  {1437} (\bibinfo {year} {2013})}\BibitemShut {NoStop}%
\bibitem [{\citenamefont {Bean}\ and\ \citenamefont
  {Livingston}(1964)}]{Bean1964}%
  \BibitemOpen
  \bibfield  {author} {\bibinfo {author} {\bibfnamefont {C.~P.}\ \bibnamefont
  {Bean}}\ and\ \bibinfo {author} {\bibfnamefont {J.~D.}\ \bibnamefont
  {Livingston}},\ }\href@noop {} {\bibfield  {journal} {\bibinfo  {journal}
  {Phys. Rev. Lett.}\ }\textbf {\bibinfo {volume} {12}},\ \bibinfo {pages} {14}
  (\bibinfo {year} {1964})}\BibitemShut {NoStop}%
\bibitem [{\citenamefont {Rogachev}\ \emph {et~al.}(2006)\citenamefont
  {Rogachev}, \citenamefont {Wei}, \citenamefont {Pekker}, \citenamefont
  {Bollinger}, \citenamefont {Goldbart},\ and\ \citenamefont
  {Bezryadin}}]{Rogachev2006}%
  \BibitemOpen
  \bibfield  {author} {\bibinfo {author} {\bibfnamefont {A.}~\bibnamefont
  {Rogachev}}, \bibinfo {author} {\bibfnamefont {T.-C.}\ \bibnamefont {Wei}},
  \bibinfo {author} {\bibfnamefont {D.}~\bibnamefont {Pekker}}, \bibinfo
  {author} {\bibfnamefont {A.~T.}\ \bibnamefont {Bollinger}}, \bibinfo {author}
  {\bibfnamefont {P.~M.}\ \bibnamefont {Goldbart}},\ and\ \bibinfo {author}
  {\bibfnamefont {A.}~\bibnamefont {Bezryadin}},\ }\href@noop {} {\bibfield
  {journal} {\bibinfo  {journal} {Phys. Rev. Lett.}\ }\textbf {\bibinfo
  {volume} {97}},\ \bibinfo {pages} {137001} (\bibinfo {year}
  {2006})}\BibitemShut {NoStop}%
\bibitem [{\citenamefont {Stan}\ \emph {et~al.}(2004)\citenamefont {Stan},
  \citenamefont {Field},\ and\ \citenamefont {Martinis}}]{Stan2004}%
  \BibitemOpen
  \bibfield  {author} {\bibinfo {author} {\bibfnamefont {G.}~\bibnamefont
  {Stan}}, \bibinfo {author} {\bibfnamefont {S.~B.}\ \bibnamefont {Field}},\
  and\ \bibinfo {author} {\bibfnamefont {J.~M.}\ \bibnamefont {Martinis}},\
  }\href@noop {} {\bibfield  {journal} {\bibinfo  {journal} {Phys. Rev. Lett.}\
  }\textbf {\bibinfo {volume} {92}},\ \bibinfo {pages} {097003} (\bibinfo
  {year} {2004})}\BibitemShut {NoStop}%
\bibitem [{\citenamefont {Likharev}(1971)}]{Likharev1971}%
  \BibitemOpen
  \bibfield  {author} {\bibinfo {author} {\bibfnamefont {K.~K.}\ \bibnamefont
  {Likharev}},\ }\href@noop {} {\bibfield  {journal} {\bibinfo  {journal}
  {Radiophys. Quantum Electron.}\ }\textbf {\bibinfo {volume} {14}},\ \bibinfo
  {pages} {722} (\bibinfo {year} {1971})}\BibitemShut {NoStop}%
\bibitem [{\citenamefont {Maksimova}(1998)}]{Maksimova1998}%
  \BibitemOpen
  \bibfield  {author} {\bibinfo {author} {\bibfnamefont {G.~M.}\ \bibnamefont
  {Maksimova}},\ }\href@noop {} {\bibfield  {journal} {\bibinfo  {journal}
  {Phys. Solid State}\ }\textbf {\bibinfo {volume} {40}},\ \bibinfo {pages}
  {1607} (\bibinfo {year} {1998})}\BibitemShut {NoStop}%
\bibitem [{\citenamefont {Berg}\ \emph {et~al.}(2007)\citenamefont {Berg},
  \citenamefont {Fradkin}, \citenamefont {Kim}, \citenamefont {Kivelson},
  \citenamefont {Oganesyan}, \citenamefont {Tranquada},\ and\ \citenamefont
  {Zhang}}]{Berg2007}%
  \BibitemOpen
  \bibfield  {author} {\bibinfo {author} {\bibfnamefont {E.}~\bibnamefont
  {Berg}}, \bibinfo {author} {\bibfnamefont {E.}~\bibnamefont {Fradkin}},
  \bibinfo {author} {\bibfnamefont {E.-A.}\ \bibnamefont {Kim}}, \bibinfo
  {author} {\bibfnamefont {S.~A.}\ \bibnamefont {Kivelson}}, \bibinfo {author}
  {\bibfnamefont {V.}~\bibnamefont {Oganesyan}}, \bibinfo {author}
  {\bibfnamefont {J.~M.}\ \bibnamefont {Tranquada}},\ and\ \bibinfo {author}
  {\bibfnamefont {S.~C.}\ \bibnamefont {Zhang}},\ }\href@noop {} {\bibfield
  {journal} {\bibinfo  {journal} {Phys. Rev. Lett.}\ }\textbf {\bibinfo
  {volume} {99}},\ \bibinfo {pages} {127003} (\bibinfo {year}
  {2007})}\BibitemShut {NoStop}%
\bibitem [{\citenamefont {Ruan}\ \emph {et~al.}(2018)\citenamefont {Ruan},
  \citenamefont {Li}, \citenamefont {Hu}, \citenamefont {Hao}, \citenamefont
  {Li}, \citenamefont {Cai}, \citenamefont {Zhou}, \citenamefont {Lee},\ and\
  \citenamefont {Wang}}]{Ruan2018}%
  \BibitemOpen
  \bibfield  {author} {\bibinfo {author} {\bibfnamefont {W.}~\bibnamefont
  {Ruan}}, \bibinfo {author} {\bibfnamefont {X.}~\bibnamefont {Li}}, \bibinfo
  {author} {\bibfnamefont {C.}~\bibnamefont {Hu}}, \bibinfo {author}
  {\bibfnamefont {Z.}~\bibnamefont {Hao}}, \bibinfo {author} {\bibfnamefont
  {H.}~\bibnamefont {Li}}, \bibinfo {author} {\bibfnamefont {P.}~\bibnamefont
  {Cai}}, \bibinfo {author} {\bibfnamefont {X.}~\bibnamefont {Zhou}}, \bibinfo
  {author} {\bibfnamefont {D.-H.}\ \bibnamefont {Lee}},\ and\ \bibinfo {author}
  {\bibfnamefont {Y.}~\bibnamefont {Wang}},\ }\href@noop {} {\bibfield
  {journal} {\bibinfo  {journal} {Nat. Phys.}\ }\textbf {\bibinfo {volume}
  {14}},\ \bibinfo {pages} {1178} (\bibinfo {year} {2018})}\BibitemShut
  {NoStop}%
\bibitem [{\citenamefont {Ye}\ \emph {et~al.}(2023)\citenamefont {Ye},
  \citenamefont {Zou}, \citenamefont {Yan}, \citenamefont {Ji}, \citenamefont
  {Xu}, \citenamefont {Dong}, \citenamefont {Chen}, \citenamefont {Zhou},\ and\
  \citenamefont {Wang}}]{Ye2023}%
  \BibitemOpen
  \bibfield  {author} {\bibinfo {author} {\bibfnamefont {S.}~\bibnamefont
  {Ye}}, \bibinfo {author} {\bibfnamefont {C.}~\bibnamefont {Zou}}, \bibinfo
  {author} {\bibfnamefont {H.}~\bibnamefont {Yan}}, \bibinfo {author}
  {\bibfnamefont {Y.}~\bibnamefont {Ji}}, \bibinfo {author} {\bibfnamefont
  {M.}~\bibnamefont {Xu}}, \bibinfo {author} {\bibfnamefont {Z.}~\bibnamefont
  {Dong}}, \bibinfo {author} {\bibfnamefont {Y.}~\bibnamefont {Chen}}, \bibinfo
  {author} {\bibfnamefont {X.}~\bibnamefont {Zhou}},\ and\ \bibinfo {author}
  {\bibfnamefont {Y.}~\bibnamefont {Wang}},\ }\href@noop {} {\bibfield
  {journal} {\bibinfo  {journal} {Nat. Phys.}\ }\textbf {\bibinfo {volume}
  {19}},\ \bibinfo {pages} {1301} (\bibinfo {year} {2023})}\BibitemShut
  {NoStop}%
\bibitem [{\citenamefont {Wollny}\ and\ \citenamefont
  {Vojta}(2009)}]{Wollny2009}%
  \BibitemOpen
  \bibfield  {author} {\bibinfo {author} {\bibfnamefont {A.}~\bibnamefont
  {Wollny}}\ and\ \bibinfo {author} {\bibfnamefont {M.}~\bibnamefont {Vojta}},\
  }\href@noop {} {\bibfield  {journal} {\bibinfo  {journal} {Phys. Rev. B}\
  }\textbf {\bibinfo {volume} {80}},\ \bibinfo {pages} {132504} (\bibinfo
  {year} {2009})}\BibitemShut {NoStop}%
\bibitem [{\citenamefont {Wilson}(2006)}]{Wilson2006}%
  \BibitemOpen
  \bibfield  {author} {\bibinfo {author} {\bibfnamefont {J.~A.}\ \bibnamefont
  {Wilson}},\ }\href@noop {} {\bibfield  {journal} {\bibinfo  {journal} {J.
  Phys.: Condens. Matter}\ }\textbf {\bibinfo {volume} {18}},\ \bibinfo {pages}
  {R69} (\bibinfo {year} {2006})}\BibitemShut {NoStop}%
\bibitem [{\citenamefont {Moshchalkov}\ \emph
  {et~al.}(2000{\natexlab{a}})\citenamefont {Moshchalkov}, \citenamefont
  {Trappeniers}, \citenamefont {Teniers}, \citenamefont {Vanacken},
  \citenamefont {Wagner},\ and\ \citenamefont
  {Bruynseraede}}]{Moshchalkov2000a}%
  \BibitemOpen
  \bibfield  {author} {\bibinfo {author} {\bibfnamefont {V.~V.}\ \bibnamefont
  {Moshchalkov}}, \bibinfo {author} {\bibfnamefont {L.}~\bibnamefont
  {Trappeniers}}, \bibinfo {author} {\bibfnamefont {G.}~\bibnamefont
  {Teniers}}, \bibinfo {author} {\bibfnamefont {J.}~\bibnamefont {Vanacken}},
  \bibinfo {author} {\bibfnamefont {P.}~\bibnamefont {Wagner}},\ and\ \bibinfo
  {author} {\bibfnamefont {Y.}~\bibnamefont {Bruynseraede}},\ }\href@noop {}
  {\bibfield  {journal} {\bibinfo  {journal} {Physica C}\ }\textbf {\bibinfo
  {volume} {341--348}},\ \bibinfo {pages} {1799} (\bibinfo {year}
  {2000}{\natexlab{a}})}\BibitemShut {NoStop}%
\bibitem [{\citenamefont {Moshchalkov}\ \emph
  {et~al.}(2000{\natexlab{b}})\citenamefont {Moshchalkov}, \citenamefont
  {Trappeniers},\ and\ \citenamefont {Vanacken}}]{Moshchalkov2000b}%
  \BibitemOpen
  \bibfield  {author} {\bibinfo {author} {\bibfnamefont {V.~V.}\ \bibnamefont
  {Moshchalkov}}, \bibinfo {author} {\bibfnamefont {L.}~\bibnamefont
  {Trappeniers}},\ and\ \bibinfo {author} {\bibfnamefont {J.}~\bibnamefont
  {Vanacken}},\ }\href@noop {} {\bibfield  {journal} {\bibinfo  {journal}
  {Physica C}\ }\textbf {\bibinfo {volume} {341--348}},\ \bibinfo {pages} {341}
  (\bibinfo {year} {2000}{\natexlab{b}})}\BibitemShut {NoStop}%
\bibitem [{\citenamefont {Li}\ \emph {et~al.}(2007)\citenamefont {Li},
  \citenamefont {H{\"u}cker}, \citenamefont {Gu}, \citenamefont {Tsvelik},\
  and\ \citenamefont {Tranquada}}]{Li2007}%
  \BibitemOpen
  \bibfield  {author} {\bibinfo {author} {\bibfnamefont {Q.}~\bibnamefont
  {Li}}, \bibinfo {author} {\bibfnamefont {M.}~\bibnamefont {H{\"u}cker}},
  \bibinfo {author} {\bibfnamefont {G.~D.}\ \bibnamefont {Gu}}, \bibinfo
  {author} {\bibfnamefont {A.~M.}\ \bibnamefont {Tsvelik}},\ and\ \bibinfo
  {author} {\bibfnamefont {J.~M.}\ \bibnamefont {Tranquada}},\ }\href@noop {}
  {\bibfield  {journal} {\bibinfo  {journal} {Phys. Rev. Lett.}\ }\textbf
  {\bibinfo {volume} {99}},\ \bibinfo {pages} {067001} (\bibinfo {year}
  {2007})}\BibitemShut {NoStop}%
\bibitem [{\citenamefont {Silotia}\ \emph {et~al.}(2024)\citenamefont
  {Silotia}, \citenamefont {Kumari}, \citenamefont {Vashist},\ and\
  \citenamefont {Chakraverty}}]{Silotia2024}%
  \BibitemOpen
  \bibfield  {author} {\bibinfo {author} {\bibfnamefont {H.}~\bibnamefont
  {Silotia}}, \bibinfo {author} {\bibfnamefont {A.}~\bibnamefont {Kumari}},
  \bibinfo {author} {\bibfnamefont {A.}~\bibnamefont {Vashist}},\ and\ \bibinfo
  {author} {\bibfnamefont {S.}~\bibnamefont {Chakraverty}},\ }\href@noop {}
  {\bibfield  {journal} {\bibinfo  {journal} {Phys. Rev. B}\ }\textbf {\bibinfo
  {volume} {109}},\ \bibinfo {pages} {245405} (\bibinfo {year}
  {2024})}\BibitemShut {NoStop}%
\bibitem [{\citenamefont {Nakamura}\ and\ \citenamefont
  {Kimura}(2009)}]{Nakamura2009}%
  \BibitemOpen
  \bibfield  {author} {\bibinfo {author} {\bibfnamefont {H.}~\bibnamefont
  {Nakamura}}\ and\ \bibinfo {author} {\bibfnamefont {T.}~\bibnamefont
  {Kimura}},\ }\href@noop {} {\bibfield  {journal} {\bibinfo  {journal} {Phys.
  Rev. B}\ }\textbf {\bibinfo {volume} {80}},\ \bibinfo {pages} {121308}
  (\bibinfo {year} {2009})}\BibitemShut {NoStop}%
\bibitem [{\citenamefont {Zou}\ \emph {et~al.}(2022)\citenamefont {Zou},
  \citenamefont {Shin}, \citenamefont {Wei}, \citenamefont {Fan}, \citenamefont
  {Davidson}, \citenamefont {Guo}, \citenamefont {Chen}, \citenamefont {Zou},\
  and\ \citenamefont {Cheng}}]{Zou2022}%
  \BibitemOpen
  \bibfield  {author} {\bibinfo {author} {\bibfnamefont {Y.}~\bibnamefont
  {Zou}}, \bibinfo {author} {\bibfnamefont {H.}~\bibnamefont {Shin}}, \bibinfo
  {author} {\bibfnamefont {H.}~\bibnamefont {Wei}}, \bibinfo {author}
  {\bibfnamefont {Y.}~\bibnamefont {Fan}}, \bibinfo {author} {\bibfnamefont
  {B.~A.}\ \bibnamefont {Davidson}}, \bibinfo {author} {\bibfnamefont {E.-J.}\
  \bibnamefont {Guo}}, \bibinfo {author} {\bibfnamefont {Q.}~\bibnamefont
  {Chen}}, \bibinfo {author} {\bibfnamefont {K.}~\bibnamefont {Zou}},\ and\
  \bibinfo {author} {\bibfnamefont {Z.~G.}\ \bibnamefont {Cheng}},\ }\href@noop
  {} {\bibfield  {journal} {\bibinfo  {journal} {npj Quantum Mater.}\ }\textbf
  {\bibinfo {volume} {7}},\ \bibinfo {pages} {122} (\bibinfo {year}
  {2022})}\BibitemShut {NoStop}%
\bibitem [{\citenamefont {Zhang}\ \emph {et~al.}(2025)\citenamefont {Zhang},
  \citenamefont {Qin}, \citenamefont {Sun}, \citenamefont {Hong},\ and\
  \citenamefont {Xie}}]{Zhang2025}%
  \BibitemOpen
  \bibfield  {author} {\bibinfo {author} {\bibfnamefont {M.}~\bibnamefont
  {Zhang}}, \bibinfo {author} {\bibfnamefont {M.}~\bibnamefont {Qin}}, \bibinfo
  {author} {\bibfnamefont {Y.}~\bibnamefont {Sun}}, \bibinfo {author}
  {\bibfnamefont {S.}~\bibnamefont {Hong}},\ and\ \bibinfo {author}
  {\bibfnamefont {Y.}~\bibnamefont {Xie}},\ }\href
  {https://arxiv.org/abs/2506.16298} {\bibinfo {title} {Universally enhanced
  superconductivity and coexisting ferroelectricity at oxide interfaces}}
  (\bibinfo {year} {2025}),\ \Eprint {https://arxiv.org/abs/2506.16298}
  {arXiv:2506.16298 [cond-mat.supr-con]} \BibitemShut {NoStop}%
\bibitem [{\citenamefont {Trybu{\l}a}\ \emph {et~al.}(2015)\citenamefont
  {Trybu{\l}a}, \citenamefont {Miga}, \citenamefont {{\L}o{\'s}}, \citenamefont
  {Trybu{\l}a},\ and\ \citenamefont {Dec}}]{Trybula2015}%
  \BibitemOpen
  \bibfield  {author} {\bibinfo {author} {\bibfnamefont {Z.}~\bibnamefont
  {Trybu{\l}a}}, \bibinfo {author} {\bibfnamefont {S.}~\bibnamefont {Miga}},
  \bibinfo {author} {\bibfnamefont {S.}~\bibnamefont {{\L}o{\'s}}}, \bibinfo
  {author} {\bibfnamefont {M.}~\bibnamefont {Trybu{\l}a}},\ and\ \bibinfo
  {author} {\bibfnamefont {J.}~\bibnamefont {Dec}},\ }\href@noop {} {\bibfield
  {journal} {\bibinfo  {journal} {Solid State Commun.}\ }\textbf {\bibinfo
  {volume} {209--210}},\ \bibinfo {pages} {23} (\bibinfo {year}
  {2015})}\BibitemShut {NoStop}%
\bibitem [{\citenamefont {Tyunina}\ \emph {et~al.}(2010)\citenamefont
  {Tyunina}, \citenamefont {Narkilahti}, \citenamefont {Plekh}, \citenamefont
  {Oja}, \citenamefont {Nieminen}, \citenamefont {Dejneka},\ and\ \citenamefont
  {Trepakov}}]{Tyunina2010}%
  \BibitemOpen
  \bibfield  {author} {\bibinfo {author} {\bibfnamefont {M.}~\bibnamefont
  {Tyunina}}, \bibinfo {author} {\bibfnamefont {J.}~\bibnamefont {Narkilahti}},
  \bibinfo {author} {\bibfnamefont {M.}~\bibnamefont {Plekh}}, \bibinfo
  {author} {\bibfnamefont {R.}~\bibnamefont {Oja}}, \bibinfo {author}
  {\bibfnamefont {R.~M.}\ \bibnamefont {Nieminen}}, \bibinfo {author}
  {\bibfnamefont {A.}~\bibnamefont {Dejneka}},\ and\ \bibinfo {author}
  {\bibfnamefont {V.}~\bibnamefont {Trepakov}},\ }\href@noop {} {\bibfield
  {journal} {\bibinfo  {journal} {Phys. Rev. Lett.}\ }\textbf {\bibinfo
  {volume} {104}},\ \bibinfo {pages} {227601} (\bibinfo {year}
  {2010})}\BibitemShut {NoStop}%
\bibitem [{\citenamefont {Junquera}\ \emph {et~al.}(2023)\citenamefont
  {Junquera}, \citenamefont {Nahas}, \citenamefont {Prokhorenko}, \citenamefont
  {Bellaiche}, \citenamefont {{\'I}{\~n}iguez}, \citenamefont {Schlom},
  \citenamefont {Chen}, \citenamefont {Salahuddin}, \citenamefont {Muller},
  \citenamefont {Martin},\ and\ \citenamefont {Ramesh}}]{Junquera2023}%
  \BibitemOpen
  \bibfield  {author} {\bibinfo {author} {\bibfnamefont {J.}~\bibnamefont
  {Junquera}}, \bibinfo {author} {\bibfnamefont {Y.}~\bibnamefont {Nahas}},
  \bibinfo {author} {\bibfnamefont {S.}~\bibnamefont {Prokhorenko}}, \bibinfo
  {author} {\bibfnamefont {L.}~\bibnamefont {Bellaiche}}, \bibinfo {author}
  {\bibfnamefont {J.}~\bibnamefont {{\'I}{\~n}iguez}}, \bibinfo {author}
  {\bibfnamefont {D.~G.}\ \bibnamefont {Schlom}}, \bibinfo {author}
  {\bibfnamefont {L.-Q.}\ \bibnamefont {Chen}}, \bibinfo {author}
  {\bibfnamefont {S.}~\bibnamefont {Salahuddin}}, \bibinfo {author}
  {\bibfnamefont {D.~A.}\ \bibnamefont {Muller}}, \bibinfo {author}
  {\bibfnamefont {L.~W.}\ \bibnamefont {Martin}},\ and\ \bibinfo {author}
  {\bibfnamefont {R.}~\bibnamefont {Ramesh}},\ }\href@noop {} {\bibfield
  {journal} {\bibinfo  {journal} {Rev. Mod. Phys.}\ }\textbf {\bibinfo {volume}
  {95}},\ \bibinfo {pages} {025001} (\bibinfo {year} {2023})}\BibitemShut
  {NoStop}%
\bibitem [{\citenamefont {Kleemann}\ \emph {et~al.}(2011)\citenamefont
  {Kleemann}, \citenamefont {Dec}, \citenamefont {Miga},\ and\ \citenamefont
  {Rytz}}]{Kleemann2011}%
  \BibitemOpen
  \bibfield  {author} {\bibinfo {author} {\bibfnamefont {W.}~\bibnamefont
  {Kleemann}}, \bibinfo {author} {\bibfnamefont {J.}~\bibnamefont {Dec}},
  \bibinfo {author} {\bibfnamefont {S.}~\bibnamefont {Miga}},\ and\ \bibinfo
  {author} {\bibfnamefont {D.}~\bibnamefont {Rytz}},\ }\href@noop {} {\bibfield
   {journal} {\bibinfo  {journal} {Z. Kristallogr.}\ }\textbf {\bibinfo
  {volume} {226}},\ \bibinfo {pages} {145} (\bibinfo {year}
  {2011})}\BibitemShut {NoStop}%
\bibitem [{\citenamefont {Venditti}\ \emph {et~al.}(2023)\citenamefont
  {Venditti}, \citenamefont {Temperini}, \citenamefont {Barone}, \citenamefont
  {Lorenzana},\ and\ \citenamefont {Gastiasoro}}]{Venditti2023}%
  \BibitemOpen
  \bibfield  {author} {\bibinfo {author} {\bibfnamefont {G.}~\bibnamefont
  {Venditti}}, \bibinfo {author} {\bibfnamefont {M.~E.}\ \bibnamefont
  {Temperini}}, \bibinfo {author} {\bibfnamefont {P.}~\bibnamefont {Barone}},
  \bibinfo {author} {\bibfnamefont {J.}~\bibnamefont {Lorenzana}},\ and\
  \bibinfo {author} {\bibfnamefont {M.~N.}\ \bibnamefont {Gastiasoro}},\
  }\href@noop {} {\bibfield  {journal} {\bibinfo  {journal} {J. Phys. Mater.}\
  }\textbf {\bibinfo {volume} {6}},\ \bibinfo {pages} {014007} (\bibinfo {year}
  {2023})}\BibitemShut {NoStop}%
\bibitem [{\citenamefont {Gastiasoro}\ \emph {et~al.}(2023)\citenamefont
  {Gastiasoro}, \citenamefont {Temperini}, \citenamefont {Barone},\ and\
  \citenamefont {Lorenzana}}]{Gastiasoro2023}%
  \BibitemOpen
  \bibfield  {author} {\bibinfo {author} {\bibfnamefont {M.~N.}\ \bibnamefont
  {Gastiasoro}}, \bibinfo {author} {\bibfnamefont {M.~E.}\ \bibnamefont
  {Temperini}}, \bibinfo {author} {\bibfnamefont {P.}~\bibnamefont {Barone}},\
  and\ \bibinfo {author} {\bibfnamefont {J.}~\bibnamefont {Lorenzana}},\
  }\href@noop {} {\bibfield  {journal} {\bibinfo  {journal} {Phys. Rev. Res.}\
  }\textbf {\bibinfo {volume} {5}},\ \bibinfo {pages} {023177} (\bibinfo {year}
  {2023})}\BibitemShut {NoStop}%
\bibitem [{\citenamefont {Chen}\ \emph {et~al.}(2018)\citenamefont {Chen},
  \citenamefont {Swartz}, \citenamefont {Yoon}, \citenamefont {Inoue},
  \citenamefont {Merz}, \citenamefont {Lu}, \citenamefont {Xie}, \citenamefont
  {Yuan}, \citenamefont {Hikita}, \citenamefont {Raghu},\ and\ \citenamefont
  {Hwang}}]{Chen2018}%
  \BibitemOpen
  \bibfield  {author} {\bibinfo {author} {\bibfnamefont {Z.}~\bibnamefont
  {Chen}}, \bibinfo {author} {\bibfnamefont {A.~G.}\ \bibnamefont {Swartz}},
  \bibinfo {author} {\bibfnamefont {H.}~\bibnamefont {Yoon}}, \bibinfo {author}
  {\bibfnamefont {H.}~\bibnamefont {Inoue}}, \bibinfo {author} {\bibfnamefont
  {T.~A.}\ \bibnamefont {Merz}}, \bibinfo {author} {\bibfnamefont
  {D.}~\bibnamefont {Lu}}, \bibinfo {author} {\bibfnamefont {Y.}~\bibnamefont
  {Xie}}, \bibinfo {author} {\bibfnamefont {H.}~\bibnamefont {Yuan}}, \bibinfo
  {author} {\bibfnamefont {Y.}~\bibnamefont {Hikita}}, \bibinfo {author}
  {\bibfnamefont {S.}~\bibnamefont {Raghu}},\ and\ \bibinfo {author}
  {\bibfnamefont {H.~Y.}\ \bibnamefont {Hwang}},\ }\href@noop {} {\bibfield
  {journal} {\bibinfo  {journal} {Nat. Commun.}\ }\textbf {\bibinfo {volume}
  {9}},\ \bibinfo {pages} {4008} (\bibinfo {year} {2018})}\BibitemShut
  {NoStop}%
\bibitem [{\citenamefont {Wang}\ \emph {et~al.}(2025)\citenamefont {Wang},
  \citenamefont {Hong}, \citenamefont {Pan}, \citenamefont {Zhou},\ and\
  \citenamefont {Xie}}]{Wang2025}%
  \BibitemOpen
  \bibfield  {author} {\bibinfo {author} {\bibfnamefont {Y.}~\bibnamefont
  {Wang}}, \bibinfo {author} {\bibfnamefont {S.}~\bibnamefont {Hong}}, \bibinfo
  {author} {\bibfnamefont {W.}~\bibnamefont {Pan}}, \bibinfo {author}
  {\bibfnamefont {Y.}~\bibnamefont {Zhou}},\ and\ \bibinfo {author}
  {\bibfnamefont {Y.}~\bibnamefont {Xie}},\ }\href@noop {} {\bibfield
  {journal} {\bibinfo  {journal} {Phys. Rev. X}\ }\textbf {\bibinfo {volume}
  {15}},\ \bibinfo {pages} {011006} (\bibinfo {year} {2025})}\BibitemShut
  {NoStop}%
\end{thebibliography}%

\end{document}